\documentclass[aps,twocolumn,preprintnumbers,amsmath,amssymb,superscriptaddress,floatfix,nofootinbib]{revtex4-1}
\usepackage{graphicx} 
\usepackage{amsmath,amssymb}
\usepackage{subfigure}
\usepackage[vcentermath]{youngtab}
\usepackage[colorlinks,linkcolor=blue]{hyperref}
\allowdisplaybreaks
\usepackage{multirow} 
\usepackage{newtxtext, newtxmath}
\usepackage{braket,bm}
\usepackage{enumitem}

\begin{document}

\title{Spectrum of $S$- and $P$-wave $cc\bar{q}\bar{q}'$ $(\bar{q},\bar{q}' = \bar{u}, \bar{d}, \bar{s})$ systems in a chiral SU(3) quark model}

\author{Du Wang}
\affiliation{School of Nuclear Science and Technology, University of Chinese Academy of Sciences, Beijing 101408, China}

\author{Ke-Rang Song}
\affiliation{School of Nuclear Science and Technology, University of Chinese Academy of Sciences, Beijing 101408, China}

\author{Wen-Ling Wang}
\email[Email: ]{wangwenling@buaa.edu.cn}
\affiliation{School of Physics, Beihang University, Beijing 100191, China}

\author{Fei Huang}
\email[Email: ]{huangfei@ucas.ac.cn}
\affiliation{School of Nuclear Science and Technology, University of Chinese Academy of Sciences, Beijing 101408, China}

\date{\today}

\begin{abstract}
Inspired by the resonance $T_{cc}^+(3875)$ recently observed by the LHCb Collaboration, we systematically explore the $S$- and $P$-wave $cc\bar{q}\bar{q}'$ $(\bar{q},\bar{q}' = \bar{u}, \bar{d}, \bar{s})$ systems in a chiral SU(3) quark model. The Hamiltonian contains the kinetic energy, the one-gluon-exchange (OGE) potential, the confinement potential, and the one-boson-exchange (OBE) potential stemming from the coupling of quark and chiral fields. The Schr\"odinger equation is solved by use of the variational method with the spacial trial wave functions chosen as Gaussian functions. It is found that the lowest state has a mass $3879$ MeV, isospin and spin-parity $IJ^P=01^+$, and quark constituent $cc\bar{u}\bar{d}$, in agreement with the experimentally observed $T_{cc}^+(3875)$. This state is approximately at the calculated $DD^\ast$ threshold, and has a root-mean-square radius about $0.48$ fm. These demonstrates that the $T_{cc}^+(3875)$ can be accommodated as a stable and compact tetraquark sate in the chiral SU(3) quark model. All the other $S$- and $P$-wave $cc\bar{q}\bar{q}'$ $(\bar{q},\bar{q}' = \bar{u}, \bar{d}, \bar{s})$ states lie about one hundred to few hundreds MeV higher than the corresponding meson-meson thresholds, and thus are not suggested to be candidates of stable and compact tetraquark states due to their fall-apart decays to two mesons.
\end{abstract}

\maketitle

\section{Introduction}

For long, we believe that hadrons can be divided into two categories, i.e. meson consists of a pair of quark and antiquark ($q\bar{q}$) and baryon consists of three quarks ($qqq$). However, in principle, the underlying theory of strong interactions -- quantum chromodynamics (QCD) -- allows the existences of any color-singlet states like tetraquarks, pentaquarks, and baryonium, etc. In experiment, a large number of candidates of exotic states have been reported in the last two decades after the observation of the $X(3872)$ by the Belle Collaboration in 2003 \cite{CDF:2003cab}. Theoretically, these candidates have aroused significant research interest, and plenty of works have been published to study the structures and properties of these possible exotic states. See Refs.~\cite{Hosaka:2016pey,Ali:2017jda,Guo:2017jvc,Esposito:2016noz,Lebed:2016hpi,Richard:2016eis,Chen:2016qju,Liu:2019zoy,Brambilla:2019esw} for recent reviews of the experimental and theoretical status of the investigations of exotic states.

Recently, a doubly charmed exotic meson $T_{cc}^+(3875)$ was announced by the LHCb Collaboration \cite{LHCb:2021vvq,LHCb:2021auc}. This state has spin-parity $J^P=1^+$ and quark constituent $cc\bar{u}\bar{d}$. Its mass is reported to be $M=3875.1$ MeV $+\delta m$ with $\delta m=-273\pm 61\pm5^{+11}_{-14}$ keV, very close to the meson-meson threshold of $DD^\ast$. It has a very small decay width, $\Gamma = 410 \pm 165\pm43^{+18}_{-38}$ keV, and is so far the longest living exotic state.

The $T_{cc}^+(3875)$ particle is of particular interest as it is really an exotic state, unlike most of the reported candidates that have the same quantum numbers as the traditional mesons or baryons and thus are not easy to be clearly identified as exotic states. In literature, the $T_{cc}^+(3875)$ state has been investigated by use of various methods, e.g., quark models \cite{Ebert:2007rn,Lu:2020rog,Wang:2022clw,Vijande:2009kj,Meng:2020knc,Deng:2021gnb}, QCD sum rules \cite{Navarra:2007yw,Du:2012wp,Agaev:2021vur}, lattice QCD \cite{Ikeda:2013vwa,Junnarkar:2018twb}, and one-boson-exchange (OBE) model \cite{Li:2012ss}, et al. Unfortunately, the conclusions from different works are not consistent, even they are done within the same method. In the quark model studies of Refs.~\cite{Ebert:2007rn,Lu:2020rog,Wang:2022clw}, no stable doubly-charmed tetraquark states were found because the lowest $cc\bar{u}\bar{d}$ states were reported to be about $60-200$ MeV higher than the thresholds of two open-charmed mesons for rearrangement decays. On the contrary, in the quark model studies of Refs.~\cite{Vijande:2009kj,Meng:2020knc,Deng:2021gnb}, opposite conclusions were drawn. In Ref.~\cite{Vijande:2009kj}, the state $cc\bar{u}\bar{d}$ with spin-parity $J^P=1^+$ was reported to be a compact tetraquark state with a mass $65$ MeV below the $DD^\ast$ threshold. In Ref.~\cite{Meng:2020knc}, the tetraquark state $cc\bar{u}\bar{d}$ with $J^P=1^+$ was found to have a mass $23$ MeV below the  $DD^\ast$ threshold. In Ref.~\cite{Deng:2021gnb}, the lowest $cc\bar{u}\bar{d}$ state was claimed to be a loosely bound $DD^\ast$ molecular state with $0.34$ MeV binding energy. In the studies by QCD sum rule, Refs.~\cite{Navarra:2007yw,Du:2012wp} reported that the calculated $cc\bar{u}\bar{d}$ state with $J^P=1^+$ are above the $DD^\ast$ threshold, while Ref.~\cite{Agaev:2021vur} claimed that the calculated mass $M=3868\pm 124$ MeV and width $\Gamma=489\pm 92$ keV for $cc\bar{u}\bar{d}$ with $J^P=1^+$ agree well with the experiment. In the lattice QCD study of Ref.~\cite{Ikeda:2013vwa}, neither bound nor resonance states were reported, while in Ref.~\cite{Junnarkar:2018twb}, a bound state with binding energy $23$ MeV was found. In the OBE model study of Ref.~\cite{Li:2012ss}, the doubly charmed tetraquark state with $J^P=1^+$ was suggested to be a $DD^\ast$ molecule.

The poor situation that diverging conclusions for the $cc\bar{u}\bar{d}$ tetraquark state were resulted from different theoretical works, regardless whether they were done with the same or different theoretical methods, urges upon us the importance and indispensable of further independent theoretical analysis with reliable model ingredients for a better understanding of the nature of the $cc\bar{u}\bar{d}$ tetraquark state. 

In our previous work of Ref.~\cite{Huang:2018rpb}, we have successfully described the energies of octet and decuplet baryon ground states, the binding energy of deuteron, the nucleon-nucleon ($NN$) scattering phase shifts and mixing parameters for angular momentum up to $J=6$ within a chiral SU(3) quark model. The Hamiltonian consists of the kinetic energy, the one-gluon-exchange (OGE) potential, the confinement potential, and the one-boson-exchange (OBE) potential stemming from the coupling of quark and chiral fields. It should be mentioned that the work of Ref.~\cite{Huang:2018rpb} was the first and so far the only one that reproduced the energies of octet and decuplet baryon ground states and the experimental data of $NN$ scattering in a rather consistent manner in a quark model. It solved the issue that the wave functions chosen for single baryons are not the solutions of the given Hamiltonian in the resonating-group-method (RGM) study of the baryon-baryon systems in constituent quark models.

In the present work, we further extend the chiral SU(3) quark model employed in Ref.~\cite{Huang:2018rpb} to explore the mass spectra of the $S$- and $P$-wave $cc\bar{q}\bar{q}'$ $(\bar{q},\bar{q}' = \bar{u},\bar{d},\bar{s})$ systems. The interactions between the light antiquarks $\bar{q}$ and $\bar{q}'$ are taken from Ref.~\cite{Huang:2018rpb}. The interactions associated with charm quark consist of the OGE potential and the confinement potential, and the corresponding parameters are fixed by a fit to the masses of known charmed baryons and mesons. The total wave functions of $cc\bar{q}\bar{q}'$ tetraquark systems are constructed as combinations of the individual wave functions in the color, spin, flavor, and coordinate spaces, where the spacial trial wave functions are chosen as Gaussian functions. The masses and eigenvectors for the $S$- and $P$-wave $cc\bar{q}\bar{q}'$ $(\bar{q},\bar{q}' = \bar{u},\bar{d},\bar{s})$ states are obtained by solving the Schr\"odinger equation via the variational method.

This paper is organized as follows. In Sec.~\ref{sec:Hamiltonian}, we introduce the Hamiltonian employed in the chiral SU(3) quark model and the determination of the model parameters. In Sec.~\ref{sec:Wave function}, we construct the spin, flavor, color, and spacial wave functions and present the relevant matrix elements of  interaction operators for the doubly charmed tetraquark states. The numeric results of the mass spectra of the $S$- and $P$-wave $cc\bar{q}\bar{q}'$ tetraquark states are shown in Sec.~\ref{sec:Results and discussion}, where some discussions are presented as well. Finally, in Sec.~\ref{sec:Summary}, we give a brief summary.

\section{Hamiltonian} \label{sec:Hamiltonian}

The Hamiltonian of the $cc\bar{q}\bar{q}'$ $(\bar{q},\bar{q}' = \bar{u}, \bar{d}, \bar{s})$ system consists of the masses of constituent quarks, the kinetic energy of the system, and the potential between each pair of the constituent quarks, e.g.,
\begin{equation}  \label{eq:Hamiltonian}
H = \sum_{i=1}^4 \left( m_i + T_i \right ) - T_G + \sum_{1=i<j}^4 \left( V_{ij}^{\text{conf}} + V_{ij}^{\text{OGE}} + V_{ij}^{\text{OBE}} \right),
\end{equation}
where $m_i$ and $T_i$ represent the mass and kinetic energy of the $i$th constituent quark, respectively, and $T_G$ denotes the kinetic energy of the center-of-mass motion of the $cc\bar{q}\bar{q}'$ system,
\begin{equation}
T_i = \frac{{\boldsymbol{p}}_i^2}{2m_i}, \qquad T_G = \frac{ \left(\sum_{i=1}^4 \boldsymbol{p}_i\right)^2  }{ 2 \sum_{i=1}^4 m_i },
\end{equation}
with $\boldsymbol{p}_i$ being the three-momentum of the $i$th constituent quark. The potential between each pair of constituent quarks is composed of three parts: the phenomenological confinement potential $V_{ij}^{\text{conf}}$, the OGE potential $ V_{ij}^{\text{OGE}}$, and the OBE potential $V_{ij}^{\text{OBE}}$ stemming from the coupling of quark and chiral fields. The last one exists only between a pair of light quarks (antiquarks) in a chiral SU(3) quark model \cite{Huang:2018rpb}. 

The confinement potential $V_{ij}^{\text{conf}}$ describes the long-range non-perturbative QCD effects. In the present work, the linear type confinement potential is adopted,
\begin{equation}
V_{ij}^{\text{conf}}=-\boldsymbol{\lambda}_i^c \cdot \boldsymbol{\lambda}_j^c \left( a_{ij} r_{ij} + a_{ij}^0 \right),   \label{Eq:Conf}
\end{equation}
where $\boldsymbol{\lambda}_i^c$ is the usual Gell-Mann matrix of the color SU(3) group, and $a_{ij}$ and $a_{ij}^0$ are parameters of the confinement strength and zero-point energy, respectively. 

The OGE potential $V_{ij}^{\text{OGE}}$ describes the short-range perturbative QCD effects. As usual, it can be written as
\begin{align}
V_{ij}^{\rm OGE} = & \; \frac{g_i g_j}{4} \boldsymbol{\lambda}_i^c \cdot \boldsymbol{\lambda}_j^c  \left[ \frac{1}{r_{ij}} - \frac{\mu_{ij}^3}{2} \frac{e^{-\mu_{ij}^2 r_{ij}^2}}{\mu_{ij} r_{ij}} \left( \frac{1}{m_i^2} + \frac{1}{m_j^2} \right. \right. \notag \\[3pt]
& \left.\left. + \, \frac{4}{3} \frac{\boldsymbol{\sigma}_i \cdot \boldsymbol{\sigma}_j}{m_i m_j}  \right)  \right] + V_{\text{ls}}^{\rm OGE}(\boldsymbol{r}_{ij}) + V_{\text{ten}}^{\rm OGE}(\boldsymbol{r}_{ij}),   \label{Eq:OGE}
\end{align}
with
\begin{align}
V^{\rm OGE}_{\rm ls}({\boldsymbol r}_{ij}) = & -\frac{g_i g_j}{4} \boldsymbol{\lambda}^c_i \cdot \boldsymbol{\lambda}^c_j \frac{m_i^2+m_j^2+4m_im_j}{8m_i^2 m_j^2}\frac{1}{r^3_{ij}}  \nonumber \\[3pt]
& \times  \left[ {\boldsymbol{L} \cdot \left( \boldsymbol{\sigma}_i + \boldsymbol{\sigma}_j \right)} \right], \\[6pt] 
V_{\text{ten}}^{\rm OGE}(\boldsymbol{r}_{ij}) = & -\frac{g_i g_j}{4} \boldsymbol{\lambda}_i^c \cdot \boldsymbol{\lambda}_j^c \frac{1}{4 m_i m_j} \frac{1}{r_{ij}^{3}} \notag\\[3pt]
& \times \left[ 3\left( \boldsymbol{\sigma}_i \cdot \hat{\boldsymbol{r}}_{ij} \right) \left( \boldsymbol{\sigma}_j \cdot \hat{\boldsymbol{r}}_{ij} \right) - \boldsymbol{\sigma}_i \cdot \boldsymbol{\sigma}_j \right],  \label{Eq:OGE-ten}
\end{align}
where $g_{i(j)}$ is the OGE coupling constant for the $i(j)$-th constituent quark, and $\mu_{ij}$ is defined as $\mu_{ij}\equiv \beta \frac{m_i m_j}{m_i+m_j}$ with $\beta$ being a model parameter. 

Note that in both confinement potential and OGE potential [cf. Eqs.~(\ref{Eq:Conf})-(\ref{Eq:OGE-ten})], the Gell-Mann matrix $\boldsymbol{\lambda}^c$ for a quark should be replaced by $-{\boldsymbol{\lambda}^c}^*$ for an antiquark.

In the chiral SU(3) quark model, the OBE potential is introduced in such a way that the Lagrangian of the  quark and chiral fields is invariant under the chiral SU(3) transition \cite{Zhang:1997ny,Huang:2004ke,Huang:2004sj,Huang:2018rpb}, which gives a natural explanation of the relatively large constituent quark masses via the mechanism of spontaneous chiral symmetry breaking. At the meanwhile, the Goldstone bosons obtain their physical masses via the obvious chiral symmetry breaking caused by the tiny current quark masses. The OBE potential provides the necessary medium- and long-range attraction in light quark systems. In the $NN$ systems, it has been shown that such attraction is rather important for describing the experimental data \cite{Huang:2018rpb}. 

The OBE potential between a pair of light quarks (antiquarks) reads 
\begin{equation}
V_{ij}^{\text{OBE}} = \sum_{a=0}^8 V_{ij}^{\sigma_a} + \sum_{a=0}^8 V_{ij}^{\pi_a},
\end{equation}
where the first and second terms represent the potential stemming from the exchanges of the scalar nonet mesons and pseudoscalar nonet mesons, respectively. The explicit expressions of $V_{ij}^{\sigma_a}$ and $V_{ij}^{\pi_a}$ are
\begin{align}
V_{ij}^{\sigma_a} = & - C(g_{\mathrm{ch}}, m'_{\sigma_a}, \Lambda) \, Y_1(m'_{\sigma_a}, \Lambda, r_{ij}) \left(\lambda_i^a \lambda_j^a\right) \notag \\[3pt]
& + V_{\mathrm{ls}}^{\sigma_a}(\boldsymbol{r}_{ij}),  \\[6pt]
V_{ij}^{\pi_a} = & ~ C(g_{\mathrm{ch}}, m'_{\pi_a}, \Lambda) \frac{m_{\pi_a}^{\prime 2} p_{ij}^a}{48} Y_3(m'_{\pi_a}, \Lambda, r_{ij}) \left(\lambda_i^a \lambda_j^a \right) \notag \\[3pt]
& \times \left( \boldsymbol{\sigma}_i \cdot \boldsymbol{\sigma}_j \right)  + V_{\mathrm{ten}}^{\pi_a}(\boldsymbol{r}_{ij}),
\end{align}
with
\begin{align}
V_{\rm ls}^{\sigma_a}(\boldsymbol{r}_{ij}) = & - C(g_{\mathrm{ch}}, m_{\sigma_a}^{\prime}, \Lambda) \frac{m_{\sigma_a}^{\prime 2} s_{ij}^a}{8} Z_3(m_{\sigma_a}^{\prime}, \Lambda, r_{ij}) \notag \\[3pt]
& \times \left[\boldsymbol{L} \cdot\left(\boldsymbol{\sigma}_{i}+\boldsymbol{\sigma}_{j}\right)\right]\left(\lambda_i^a \lambda_j^a\right),  \\[6pt]
V_{\rm ten}^{\pi_a}(\boldsymbol{r}_{ij}) = & ~ C(g_{\mathrm{ch}}, m_{\pi_a}^{\prime}, \Lambda) \frac{m_{\pi_a}^{\prime 2} p_{ij}^a}{48} H_3(m_{\pi_a}^{\prime}, \Lambda, r_{ij}) \notag \\[3pt]
& \times \left[3\left(\boldsymbol{\sigma}_i \cdot \hat{\boldsymbol{r}}_{ij}\right)\left(\boldsymbol{\sigma}_j \cdot \hat{\boldsymbol{r}}_{ij}\right)-\boldsymbol{\sigma}_i \cdot \boldsymbol{\sigma}_j\right]\left(\lambda_i^a \lambda_j^a\right),
\end{align}
where
\begin{align}
C(g_{\mathrm{ch}}, m, \Lambda) = & ~ \frac{g_{\mathrm{ch}}^2}{4 \pi} \frac{\Lambda^2}{\Lambda^2-m^2} m,  \\[6pt]
Y_1(m, \Lambda, r) = & ~ Y(m r) - \frac{\Lambda}{m} Y(\Lambda r), \\[6pt]
Y_3(m, \Lambda, r) = & ~ Y(m r) - \left(\frac{\Lambda}{m}\right)^3 Y(\Lambda r), \\[6pt]
Z_3(m, \Lambda, r) = & ~ Z(m r) - \left(\frac{\Lambda}{m}\right)^3 Z(\Lambda r), \\[6pt]
H_3(m, \Lambda, r) = & ~  H(m r) - \left(\frac{\Lambda}{m}\right)^3 H(\Lambda r),\\[6pt]
Y(x) = & ~ \frac{1}{x} e^{-x}, \\[6pt]
Z(x) = & ~ \left(\frac{1}{x}+\frac{1}{x^{2}}\right) Y(x), \\[6pt]
H(x) = & ~ \left(1+\frac{3}{x}+\frac{3}{x^{2}}\right) Y(x),
\end{align}
and
\begin{align}
s_{ij}^a = & \left\{ \begin{array}{lccl} \dfrac{1}{m_i^2}+\dfrac{1}{m_j^2}, &&& ~(a=0,1,2,3,8) \\[12pt] \dfrac{2}{m_i m_j}, &&& ~(a=4,5,6,7) \end{array}\right.   \\[6pt]
p_{ij}^a = & \left\{ \begin{array}{lcl} \dfrac{4}{m_i m_j}, && (a=0,1,2,3,8) \\[9pt] \dfrac{\left(m_i+m_j\right)^2}{m_i^2 m_j^2}, && (a=4,5,6,7) \end{array} \right.  \\[6pt]
m_{\sigma_a}^\prime = & \left\{ \begin{array}{ll} m_{\sigma_a}, & (a=0,1,2,3,8) \\[6pt]  \sqrt{m_{\sigma_a}^2 - \left( m_i - m_j \right)^2}, & (a=4,5,6,7)  \end{array} \right. \\[6pt] 
m_{\pi_a}^\prime = & \left\{ \begin{array}{ll} m_{\pi_a}, & (a=0,1,2,3,8) \\[6pt] \sqrt{m_{\pi_a}^2 - \left( m_i - m_j \right)^2}. & (a=4,5,6,7) \end{array} \right. 
\end{align}
Here, $g_{\rm ch}$ is the quark and chiral field coupling constant, $\Lambda$ is the cutoff parameter indicating the chiral symmetry breaking scale, and $m_{\pi_a}$ and $m_{\sigma_a}$ $(a=0,1,2,\cdots,8)$ represent the masses of nonet pseudoscalar and nonet scalar mesons, respectively. 

For pseudoscalar meson exchanges, the couplings of $\eta_0$ and $\eta_8$ are considered to give the physical $\eta$ and $\eta'$ states:
\begin{align}
\left\{ \begin{array}{l} \eta = \eta_8 \cos\theta - \eta_0 \sin\theta,  \\[6pt]  \eta^\prime = \eta_8 \sin\theta + \eta_0 \cos\theta,  \end{array} \right.
\end{align}
with the mixing angle $\theta$ taken to be the empirical value $\theta=-23^\circ$.

\begin{table}[tbp]
\caption{Model parameters associated with heavy quarks. The charm quark mass $m_c$ and the parameters of zero point energies $a_{cc}^0$, $a_{cu}^0$, and $a_{cs}^0$ are in MeV. The strengths of confinement $a_{cc}$, $a_{cu}$, and $a_{cs}$ are in MeV/fm. }
\begin{tabular*}{\columnwidth}{@{\extracolsep{\fill}}cccccccc}
\hline\hline
$m_c$ & $g_c$ & $a_{cc}$ & $a_{cu}$ & $a_{cs}$ & $a_{cc}^0$ & $a_{cu}^0$ & $a_{cs}^0$  \\ \hline
$1500$ & $0.635$  &  $183.8$  &  $160.9$  &  $130.6$  &  $-61.9$  &  $-95.3$  &  $-53.6$  \\
\hline\hline
\end{tabular*}   \label{tab:parameters}
\end{table}

\begin{table}[tbp]
\caption{Masses (in MeV) of charmed mesons and baryons calculated by use of the parameters listed in Table~\ref{tab:parameters}. The corresponding masses from PDG \cite{ParticleDataGroup:2022pth} are listed in the last column.}
\begin{tabular*}{\columnwidth}{@{\extracolsep{\fill}}lccc}
 \hline\hline
    Particles  &  $IJ^P$  &  Masses  &  PDG values  \\
    \hline
    $D^0$        &  $\frac{1}{2}0^-$  &  $1866.9$   &  $1864.84\pm0.05$   \\
    $D^{*+}$    &   $\frac{1}{2}1^-$  &  $2011.6$   & $2010.26\pm0.05$   \\ 
    $D_s^\pm$   &  $00^-$     & $1968.4$   & $1968.35\pm0.07$   \\
    $D_s^{*\pm}$    &  $01^-$   & $2133.3$   & $2112.2\pm0.4$   \\
    $\eta_c(1S)$    &  $00^-$    & $2975.9$   & $2983.9\pm0.4$   \\ 
    $J/\psi(1S)$    &  $01^-$     &  $3096.8$   & $3096.9\pm0.006$   \\
    $\eta_c(2S)$    &  $00^-$    &  $3613.4$   &  $3638\pm 1$   \\
    $J/\psi(2S)$    &  $01^-$     &  $3686.1$   &  $3686\pm 0.01$   \\
    $\Lambda_c^+$       &  $0\frac{1}{2}^+$   & $2245.4$   & $2286.46\pm 0.14$   \\
    $\Sigma_c(2455)$    &  $1\frac{1}{2}^+$   & $2445.4$   & $2453.97\pm 0.14$   \\
    $\Sigma_c(2520)$    &  $1\frac{3}{2}^+$   & $2517.6$   & $2518.41^{+ 0.22}_{-0.18}$   \\
    $\Xi_c^+$           &  $\frac{1}{2}\frac{1}{2}^+$ & $2456.0$   & $2467.71\pm 0.23$   \\
    $\Xi_c^{'+}$        &  $\frac{1}{2}\frac{1}{2}^+$ & $2566.7$   & $2578.2\pm 0.5$   \\
    $\Xi_c(2645)^+$     &  $\frac{1}{2}\frac{3}{2}^+$ & $2642.6$   & $2645.10\pm 0.30$   \\
    $\Omega_c^0$      &  $0\frac{1}{2}^+$          & $2680.8$   & $2695.2\pm 1.7$   \\
    $\Omega_c(2770)^0$  &  $0\frac{3}{2}^+$    &  $2764.3$   & $2765.9\pm 2.0$   \\
\hline\hline
\end{tabular*}     \label{tab:fit}
\end{table}

We mention that the OBE potential between a pair of light quark and antiquark is related to that between a pair of light quarks by a $G$-parity of the exchanged meson \cite{Huang:2003we,Huang:2004sj,Wang:2007kb}. In the present work, the OBE potential only exists between a pair of light antiquarks, which is the same as that between a pair of light quarks.

The model parameters for light quarks (antiquarks) are taken from our previous work of Ref.~\cite{Huang:2018rpb}, which for the first time gave a rather consistent description in the quark model of the masses of octet and decuplet baryon ground states, the binding energy of deuteron, and the $NN$ scattering phase shifts and mixing parameters for partial waves up to total angular momentum $J=6$. These parameters are: $m_u = m_d =313$ MeV, $m_s = 470$ MeV, $m_{\sigma'} = m_\kappa = m_\epsilon = 980$ MeV, $m_\sigma = 569$ MeV, $m_\pi = 138$ MeV, $m_K = 495$ MeV, $m_\eta = 549$ MeV, $m_{\eta'} = 957$ MeV, $\Lambda = 1100$ MeV, $a_{uu}=58.39$ MeV/fm, $a_{us}=69.77$ MeV/fm, $a_{ss}=95.23$ MeV/fm, $a_{uu}^0=-24.53$ MeV, $a_{us}^0=-22.86$ MeV, $a_{ss}^0=-26.10$ MeV, $g_{u(s)}=1.164$, and $\beta=1.606$. Note that the values of some of these parameters are a little bit different from those of Ref.~\cite{Huang:2018rpb}, because the $\delta$-function in OGE potential in Ref.~\cite{Huang:2018rpb} has now been replaced by 
\begin{align}
\delta(\boldsymbol{r}_{ij}) \quad \to \quad \frac{\mu_{ij}^3}{\pi} \frac{e^{-\mu_{ij}^2 r_{ij}^2}}{\mu r_{ij}}
\end{align}
with 
\begin{align} 
\mu_{ij} = \beta \frac{m_i m_j}{m_i+m_j}
\end{align}
in the present work to avoid the problem of collapsing ground state caused by the fact that the $\delta$-potential is more attractive than $1/r^2$ and so overpowers the kinetic energy $p^2/(2m)$ for a pair of scalar quarks (antiquarks). After this replacement, the energies of octet and decuplet baryon ground states, the binding energy of deuteron, and the $NN$ scattering phase shifts and mixing parameters up to total angular momentum $J=6$ are refitted. The resulted fitting quality is almost the same as that in Ref.~\cite{Huang:2018rpb} by use of the above-mentioned parameter values.

The other model parameters are those associated with heavy charm quarks. Apart from the charm quark mass $m_c$, they are $g_c$, $a_{cu}$, $a_{cs}$, $a_{cc}$, $a_{cu}^0$, $a_{cs}^0$, and $a_{cc}^0$ from OGE and confinement potentials. All these parameters are determined by fitting the masses of known charmed mesons and baryons. The fitted values of these parameters are listed in Table~\ref{tab:parameters}, and the masses of charmed mesons and baryons calculated by use of these parameters are presented in Table~\ref{tab:fit}, where the experimental values \cite{ParticleDataGroup:2022pth} are also listed for comparison. One sees that the calculated masses of charmed mesons and baryons are rather close to their experimental values.

We mention that although the chiral SU(3) quark model offers a natural explanation for the relatively large constituent quark masses and reasonably reproduces the energies of individual baryons as well as the $NN$ and $YN$ ($Y=\Lambda,\Sigma$) scattering data, its success remains incompletely understood within the QCD framework.

\section{Wave functions} \label{sec:Wave function}

\begin{table}[tbp]
\caption{Color matrix elements of $\boldsymbol{\lambda}_i^c \cdot \boldsymbol{\lambda}_j^c$.}
\begin{tabular*}{\columnwidth}{@{\extracolsep{\fill}}lcccccc}
\hline\hline
   & $\boldsymbol{\lambda}_1^c \cdot \boldsymbol{\lambda}_2^c$ & $\boldsymbol{\lambda}_3^c \cdot \boldsymbol{\lambda}_4^c$ & $\boldsymbol{\lambda}_1^c \cdot \boldsymbol{\lambda}_3^c$ & $\boldsymbol{\lambda}_2^c \cdot \boldsymbol{\lambda}_4^c$  & $\boldsymbol{\lambda}_1^c \cdot \boldsymbol{\lambda}_4^c$ & $\boldsymbol{\lambda}_2^c \cdot \boldsymbol{\lambda}_3^c$  \\[3pt]
\hline
$\langle 6\bar{6} | \hat{O} | 6\bar{6}\rangle$ & $4/3$ & $4/3$ & $-10/3$ & $-10/3$ & $-10/3$ & $-10/3$ \\
$\langle 6\bar{6} | \hat{O} | \bar{3}3\rangle$ & $0$ & $0$ & $-2\sqrt{2}$ & $-2\sqrt{2}$ & $2\sqrt{2}$ & $2\sqrt{2}$ \\
$\langle \bar{3}3 | \hat{O} | \bar{3}3\rangle$ & $-8/3$  & $-8/3$ & $-4/3$ & $-4/3$ & $-4/3$ & $-4/3$ \\
\hline\hline
\end{tabular*}   \label{tab:color_matrix_element}
\end{table}

\begin{table}[tbp]
\caption{Flavor matrix elements of $1$, $\sum_{a=1}^3 \lambda_{\bar{q}}^a \lambda_{\bar{q}'}^a$, $\sum_{a=4}^7 \lambda_{\bar{q}}^a \lambda_{\bar{q}'}^a$, and $\lambda_{\bar{q}}^8 \lambda_{\bar{q}'}^8$.}
\begin{tabular*}{\columnwidth}{@{\extracolsep{\fill}}cccccc}
\hline\hline
    & $[\bar{n} \bar{n}']$ & $\{\bar{n}\bar{n}'\}$ & $[\bar{n}\bar{s}]$ & $\{\bar{n} \bar{s}\}$ & $\{\bar{s} \bar{s}\}$ \\
    \hline
 $\langle 1 \rangle$ & $1$ & $1$ & $1$ & $1$ & $1$  \\
 $\langle \sum_{a=1}^3 \lambda_{\bar{q}}^a \lambda_{\bar{q}'}^a \rangle$ & $-3$ & $1$ & $0$ & $0$ & $0$  \\
 $\langle \sum_{a=4}^7 \lambda_{\bar{q}}^a \lambda_{\bar{q}'}^a \rangle$ & $0$ & $0$ & $-2$ & $2$ & $0$ \\
 $\langle \lambda_{\bar{q}}^8 \lambda_{\bar{q}'}^8 \rangle$ & $1/3$ & $1/3$ & $-2/3$ & $-2/3$ & $4/3$  \\
\hline\hline
\end{tabular*}  \label{tab:flavout matrix element}
\end{table}

\begin{table}[tbp]
\caption{Spin matrix elements of $\boldsymbol{\sigma}_i \cdot \boldsymbol{\sigma}_j$.}
\begin{tabular*}{\columnwidth}{@{\extracolsep{\fill}}lcccccc}
\hline \hline
    & $\boldsymbol{\sigma}_1 \cdot \boldsymbol{\sigma}_2$  & $\boldsymbol{\sigma}_3 \cdot \boldsymbol{\sigma}_4$  & $\boldsymbol{\sigma}_1 \cdot \boldsymbol{\sigma}_3$  & $\boldsymbol{\sigma}_2 \cdot \boldsymbol{\sigma}_4$ & $\boldsymbol{\sigma}_1 \cdot \boldsymbol{\sigma}_4$  &  $\boldsymbol{\sigma}_2 \cdot \boldsymbol{\sigma}_3$  \\  \hline
 $\langle\chi_0^{00}|\hat{O}| \chi_0^{00}\rangle$ & $-3$ & $-3$ & $0$ & $0$ & $0$ & $0$  \\
 $\langle\chi_0^{11}|\hat{O}| \chi_0^{11}\rangle$ & $1$ & $1$ & $-2$ & $-2$ & $-2$ & $-2$ \\
 $\langle\chi_0^{00}|\hat{O}| \chi_0^{11}\rangle$ & $0$ & $0$ & $-\sqrt{3}$ & $-\sqrt{3}$ & $\sqrt{3}$ & $\sqrt{3}$ \\
 $\langle\chi_1^{01}|\hat{O}| \chi_1^{01}\rangle$ & $-3$ & $1$ & $0$ & $0$ & $0$ & $0$ \\
 $\langle\chi_1^{10}|\hat{O}| \chi_1^{10}\rangle$ & $1$ & $-3$ & $0$ & $0$ & $0$ & $0$ \\
 $\langle\chi_1^{11}|\hat{O}| \chi_1^{11}\rangle$ & $1$ & $1$ & $-1$ & $-1$ & $-1$ & $-1$ \\
 $\langle\chi_1^{01}|\hat{O}| \chi_1^{10}\rangle$ & $0$ & $0$ & $1$ & $1$ & $-1$ & $-1$ \\
 $\langle\chi_1^{01}|\hat{O}| \chi_1^{11}\rangle$ & $0$ & $0$ & $-\sqrt{2}$ & $\sqrt{2}$ & $-\sqrt{2}$ & $\sqrt{2}$ \\
 $\langle\chi_1^{10}|\hat{O}| \chi_1^{11}\rangle$ & $0$ & $0$ & $\sqrt{2}$ & $-\sqrt{2}$ & $-\sqrt{2}$ & $\sqrt{2}$ \\
 $\langle\chi_2^{11}|\hat{O}| \chi_2^{11}\rangle$ & $1$ & $1$ & $1$ & $1$ & $1$ & $1$ \\
\hline\hline
\end{tabular*}     \label{tab:spin matrix element}
\end{table}

\begin{table}[htb]
\caption{Wave functions of the $S$- and $P$-wave tetraquark states $cc\bar{q}\bar{q}'$ $(\bar{q},\bar{q}' = \bar{u}, \bar{d}, \bar{s})$. The superscript and subscript represent the color SU(3) representation and the spin, respectively.}
 \begin{tabular*}{\columnwidth}{@{\extracolsep{\fill}}cll}
 \hline\hline 
 $J^P$ & \multicolumn{2}{c}{Configuration} \\ 
\hline  
$0^+$ & $\big(\{cc\}_1^{\bar{3}_c}\{\bar{q}\bar{q}'\}_1^{3_c}\big)_{S=0}^{1_c}$ & $\big(\{cc\}_0^{6_c}\{\bar{q}\bar{q}'\}_0^{\bar{6}_c}\big)_{S=0}^{1_c}$ \vspace{1em} \\
$1^+$ & $\big(\{cc\}_1^{\bar{3}_c}[\bar{q}\bar{q}']_0^{3_c}\big)_{S=1}^{1_c}$ & $\big(\{cc\}_0^{6_c}[\bar{q}\bar{q}']_1^{\bar{6}_c}\big)_{S=1}^{1_c}$ \vspace{0.2em}  \\
         & $\big(\{cc\}_1^{\bar{3}_c}\{\bar{q}\bar{q}'\}_1^{3_c}\big)_{S=1}^{1_c}$ &  \vspace{1em} \\
 $2^+$ & $\big(\{cc\}_1^{\bar{3}_c}\{\bar{q}\bar{q}'\}_1^{3_c}\big)_{S=2}^{1_c}$ &  \vspace{1em} \\
 $0^-$ & $\big({\{cc\}}_{1}^{\bar{3}_c}[\bar{q}\bar{q}',\rho]_{1}^{3_c}\big)^{1_c}_{S=1}$ & $\big(\{cc,\rho\}_{1}^{6_c}{[\bar{q}\bar{q}']}_{1}^{\bar{6}_c}\big)^{1_c}_{S=1}$ \vspace{0.2em} \\
         & $\big({\{cc\}}_{1}^{\bar{3}_c}{[\bar{q}\bar{q}']}_{0}^{3_c},\lambda\big)^{1_c}_{S=1}$ & $\big({\{cc\}}_{0}^{6_c}{[\bar{q}\bar{q}']}_{1}^{\bar{6}_c},\lambda\big)^{1_c}_{S=1}$ \vspace{0.2em} \\
         & $\big(\{cc,\rho\}_{0}^{\bar{3}_c}{\{\bar{q}\bar{q}'\}}_{1}^{3_c}\big)^{1_c}_{S=1}$ & $\big(\{cc,\rho\}^{6_c}_{1}{\{\bar{q}\bar{q}'\}}^{\bar{6}_c}_{0}\big)^{1_c}_{S=1}$ \vspace{0.2em} \\
         & $\big({\{cc\}}_{1}^{\bar{3}_c}\{\bar{q}\bar{q}',\rho\}_{0}^{3_c}\big)^{1_c}_{S=1}$  &$\big({\{cc\}}_{0}^{6_c}\{\bar{q}\bar{q}',\rho\}_{1}^{\bar{6}_c}\big)^{1_c}_{S=1}$  \vspace{0.2em} \\
       & $\big({\{cc\}}_{1}^{\bar{3}_c}{\{\bar{q}\bar{q}'\}}_{1}^{3_c},\lambda\big)^{1_c}_{S=1}$ & \vspace{1em} \\
$1^-$ & $\big(\{cc,\rho\}_{0}^{\bar{3}_c}{[\bar{q}\bar{q}']}_{0}^{3_c}\big)^{1_c}_{S=0}$ & $\big(\{cc,\rho\}_{1}^{6_c}{[\bar{q}\bar{q}']}_{1}^{\bar{6}_c}\big)^{1_c}_{S=0}$  \vspace{0.2em}  \\
         & $\big({\{cc\}}_{1}^{\bar{3}_c}[\bar{q}\bar{q}',\rho]_{1}^{3_c}\big)^{1_c}_{S=0}$ & $\big(\{cc,\rho\}_{1}^{6_c}{[\bar{q}\bar{q}']}_{1}^{\bar{6}_c}\big)^{1_c}_{S=1}$ \vspace{0.2em} \\
         & $\big({\{cc\}}_{1}^{\bar{3}_c}[\bar{q}\bar{q}',\rho]_{1}^{3_c}\big)^{1_c}_{S=1}$ & $\big(\{cc,\rho\}_{1}^{6_c}{[\bar{q}\bar{q}']}_{1}^{\bar{6}_c}\big)^{1_c}_{S=2}$ \vspace{0.2em} \\
         & $\big({\{cc\}}_{1}^{\bar{3}_c}[\bar{q}\bar{q}',\rho]_{1}^{3_c}\big)^{1_c}_{S=2}$  & $\big({\{cc\}}_{0}^{6_c}[\bar{q}\bar{q}',\rho]_{0}^{\bar{6}_c}\big)^{1_c}_{S=0}$ \vspace{0.2em} \\
         & $\big({\{cc\}}_{1}^{\bar{3}_c}{[\bar{q}\bar{q}']}_{0}^{3_c},\lambda\big)^{1_c}_{S=1}$ & $\big({\{cc\}}_{0}^{6_c}{[\bar{q}\bar{q}']}_{1}^{\bar{6}_c},\lambda\big)^{1_c}_{S=1}$ \vspace{0.2em} \\
         & $\big(\{cc,\rho\}_{0}^{\bar{3}_c}{\{\bar{q}\bar{q}'\}}_{1}^{3_c}\big)^{1_c}_{S=1}$ & $\big(\{cc,\rho\}^{6_c}_{1}{\{\bar{q}\bar{q}'\}}^{\bar{6}_c}_{0}\big)^{1_c}_{S=1}$ \vspace{0.2em} \\
         & $\big({\{cc\}}_{1}^{\bar{3}_c}\{\bar{q}\bar{q}',\rho\}_{0}^{3_c}\big)^{1_c}_{S=1}$ & $\big({\{cc\}}_{0}^{6_c}\{\bar{q}\bar{q}',\rho\}_{1}^{\bar{6}_c}\big)^{1_c}_{S=1}$ \vspace{0.2em} \\
         & $\big({\{cc\}}_{1}^{\bar{3}_c}{\{\bar{q}\bar{q}'\}}_{1}^{3_c},\lambda\big)^{1_c}_{S=0}$ &$\big({\{cc\}}_{0}^{6_c}{\{\bar{q}\bar{q}'\}}_{0}^{\bar{6}_c},\lambda\big)^{1_c}_{S=0}$ \vspace{0.2em} \\
     & $\big({\{cc\}}_{1}^{\bar{3}_c}{\{\bar{q}\bar{q}'\}}_{1}^{3_c},\lambda\big)^{1_c}_{S=1}$ & \vspace{0.2em} \\
      & $\big({\{cc\}}_{1}^{\bar{3}_c}{\{\bar{q}\bar{q}'\}}_{1}^{3_c},\lambda\big)^{1_c}_{S=2}$ & \vspace{1em} \\
$2^-$ & $\big({\{cc\}}_{1}^{\bar{3}_c}[\bar{q}\bar{q}',\rho]_{1}^{3_c}\big)^{1_c}_{S=1}$ &$\big(\{cc,\rho\}_{1}^{6_c}{[\bar{q}\bar{q}']}_{1}^{\bar{6}_c}\big)^{1_c}_{S=1}$ \vspace{0.2em} \\
         & $\big({\{cc\}}_{1}^{\bar{3}_c}[\bar{q}\bar{q}',\rho]_{1}^{3_c}\big)^{1_c}_{S=2}$ &$\big(\{cc,\rho\}_{1}^{6_c}{[\bar{q}\bar{q}']}_{1}^{\bar{6}_c}\big)^{1_c}_{S=2}$ \vspace{0.2em} \\
         & $\big({\{cc\}}_{1}^{\bar{3}_c}{[\bar{q}\bar{q}']}_{0}^{3_c},\lambda\big)^{1_c}_{S=1}$ & $\big({\{cc\}}_{0}^{6_c}{[\bar{q}\bar{q}']}_{1}^{\bar{6}_c},\lambda\big)^{1_c}_{S=1}$ \vspace{0.2em} \\
         & $\big(\{cc,\rho\}_{0}^{\bar{3}_c}{\{\bar{q}\bar{q}'\}}_{1}^{3_c}\big)^{1_c}_{S=1}$ &$\big(\{cc,\rho\}^{6_c}_{1}{\{\bar{q}\bar{q}'\}}^{\bar{6}_c}_{0}\big)^{1_c}_{S=1}$ \vspace{0.2em} \\
         & $\big({\{cc\}}_{1}^{\bar{3}_c}\{\bar{q}\bar{q}',\rho\}_{0}^{3_c}\big)^{1_c}_{S=1}$ &$\big({\{cc\}}_{0}^{6_c}\{\bar{q}\bar{q}',\rho\}_{1}^{\bar{6}_c}\big)^{1_c}_{S=1}$ \vspace{0.2em} \\
    & $\big({\{cc\}}_{1}^{\bar{3}_c}{\{\bar{q}\bar{q}'\}}_{1}^{3_c},\lambda\big)^{1_c}_{S=1}$ & \vspace{0.2em}\\
    & $\big({\{cc\}}_{1}^{\bar{3}_c}{\{\bar{q}\bar{q}'\}}_{1}^{3_c},\lambda\big)^{1_c}_{S=2}$ & \vspace{1em} \\
$3^-$ & $\big({\{cc\}}_{1}^{\bar{3}_c}[\bar{q}\bar{q}',\rho]_{1}^{3_c}\big)^{1_c}_{S=2}$ & $\big(\{cc,\rho\}_{1}^{6_c}{[\bar{q}\bar{q}']}_{1}^{\bar{6}_c}\big)^{1_c}_{S=2}$ \vspace{0.2em} \\
         &$\big({\{cc\}}_{1}^{\bar{3}_c}{\{\bar{q}\bar{q}'\}}_{1}^{3_c},\lambda\big)^{1_c}_{S=2}$ & \\
\hline\hline
\end{tabular*}    \label{tab:wave function}
\end{table}

In this section, we briefly present the process of constructing the wave functions for the $S$- and $P$-wave $cc\bar{q}\bar{q}'$ $(\bar{q},\bar{q}' = \bar{u}, \bar{d}, \bar{s})$ systems. The total wave function of a tetraquark state consists of four parts, i.e. the color, flavor, spin, and space parts. 

In the color space, the exact symmetry is SU(3)$_c$. As a quark belongs to a ${\bf{3}}_c$ representation and an antiquark belongs to a ${\bf{\bar{3}}}_c$ representation, two quarks couple into either ${\bf{6}}_c$ or ${\bf{\bar{3}}}_c$ representations, i.e. ${\bf{3}}_c \otimes {\bf{3}}_c = {\bf{6}}_c \oplus {\bf{\bar{3}}}_c$, and two antiquarks couple into either ${\bf{3}}_c$ or ${\bf{\bar{6}}}_c$ representations, i.e. ${\bf{\bar{3}}}_c \otimes {\bf{\bar{3}}}_c = {\bf{3}}_c \oplus {\bf{\bar{6}}}_c$. Then the color wave function for a tetraquark state can be constructed as either $ \big[ \left( {\bf{3}}_c \otimes {\bf{3}}_c \right)_{{\bf{6}}_c} \otimes \left( {\bf{\bar{3}}}_c \otimes {\bf{\bar{3}}}_c \right)_{{\bf{\bar{6}}}_c} \big]_{{\bf{1}}_c} $ or $ \big[ \left( {\bf{3}}_c \otimes {\bf{3}}_c \right)_{{\bf{\bar{3}}}_c} \otimes \left( {\bf{\bar{3}}}_c \otimes {\bf{\bar{3}}}_c \right)_{{\bf{3}}_c} \big]_{{\bf{1}}_c} $, which can be illustrated by
\begin{equation*}
\left [ \, \Big ( ~ \yng(1) ~ \otimes ~ \yng(1) ~ \Big )_{~ \scriptsize \yng(2)} ~ \otimes ~ \left ( ~ \yng(1,1) ~ \otimes ~ \yng(1,1) ~ \right )_{~ \scriptsize \yng(2,2)} \; \right ]_{~ \scriptsize \yng(2,2,2)}
\end{equation*}
and
\begin{equation*}
\left [ \, \Big ( ~ \yng(1) ~ \otimes ~ \yng(1) ~ \Big )_{~ \scriptsize \yng(1,1)} ~ \otimes ~ \left ( ~ \yng(1,1) ~ \otimes ~ \yng(1,1) ~ \right )_{~ \scriptsize \yng(2,1,1)} \; \right ]_{~ \scriptsize \yng(2,2,2)}
\end{equation*}
respectively.
In the following parts of the paper, these two types of color wave functions are denoted as
\begin{align}
\left | 6 \bar{6} \right \rangle & \equiv \left [ {\left(cc\right)}^{\bf{6}_c} {\left(\bar{q}\bar{q}'\right)}^{{\bf\bar{6}}_c} \right ] ^{{\bf{1}}_c},  \\[6pt]
\left | \bar{3} 3 \right \rangle & \equiv \left [ {\left(cc\right)}^{{\bf{\bar{3}}}_c} {\left(\bar{q}\bar{q}'\right)}^{{\bf{3}}_c} \right ] ^{{\bf{1}}_c}.
\end{align}
By using the color wave functions $\left | 6 \bar{6} \right \rangle$ and $\left | \bar{3} 3 \right \rangle$, one can evaluate the matrix elements of $\boldsymbol{\lambda}_i^c \cdot \boldsymbol{\lambda}_j^c$ straightforwardly. The results are presented in Table~\ref{tab:color_matrix_element}.

In the flavor space, one only needs to consider the symmetry of two light antiquarks for the $cc\bar{q}\bar{q}'$ $(\bar{q},\bar{q}' = \bar{u}, \bar{d}, \bar{s})$ systems. In this paper, we use the curly brace $\{~\}$ and square bracket $[~]$ to represent the symmetric and antisymmetric flavor wave functions of two light antiquarks, i.e. $\left\{ \bar{q}\bar{q}' \right\} = \left( \bar{q} \bar{q}' +\bar{q}'\bar{q} \right)/\sqrt{2}$ and $\left[ \bar{q}\bar{q}' \right] = \left( \bar{q}\bar{q}' - \bar{q}'\bar{q} \right)/\sqrt{2}$. As the chiral fields couple only to the light quarks (antiquarks), one needs to calculate the matrix elements of the operators $1$, $\sum_{a=1}^3 \lambda_{\bar{q}}^a \lambda_{\bar{q}'}^a$, $\sum_{a=4}^7 \lambda_{\bar{q}}^a \lambda_{\bar{q}'}^a$, and $\lambda_{\bar{q}}^8 \lambda_{\bar{q}'}^8$ in the flavor space. The evaluated values are listed in Table~\ref{tab:flavout matrix element}, where $n$ and $s$ represent the non-strange and strange quarks, respectively.

In the spin space, the wave function for a $cc\bar{q}\bar{q}'$ $(\bar{q},\bar{q}' = \bar{u}, \bar{d}, \bar{s})$ system is constructed as
\begin{equation}
\chi^{S_{12} S_{34}}_S \equiv {\left| {\left(cc\right)}_{S_{12}} {\left(\bar{q}\bar{q}'\right)}_{S_{34}} \right\rangle}_S,
\end{equation}
where $S_{12}$ and $S_{34}$ represent the spin of two charm quarks and two light antiquarks, respectively, and $S$ represents the spin of the total tetraquark system. The matrix elements of the operator $\boldsymbol{\sigma}_i \cdot \boldsymbol{\sigma}_j$ evaluated based on this spin wave function are listed in Table~\ref{tab:spin matrix element}.

In the coordinate space, we define the following Jacobi coordinates for the $cc\bar{q}\bar{q}'$ $(\bar{q},\bar{q}' = \bar{u}, \bar{d}, \bar{s})$ systems:
\begin{align}
 \bm{\xi}_1 &= \bm{r}_1 - \bm{r}_2,  \\[6pt]
 \bm{\xi}_2 &= \bm{r}_3 - \bm{r}_4,  \\[6pt]
 \bm{\xi}_3 &= \frac{\bm{r}_1 + \bm{r}_2}{2} - \frac{m_3 \bm{r}_3 + m_4 \bm{r}_4}{m_3 + m_4},
\end{align}
where $\bm{r}_1$ and $\bm{r}_2$ are coordinates of two charm quarks, and $\bm{r}_3$ and $\bm{r}_4$ are coordinates of two light antiquarks with $m_3$ and $m_4$ being their masses, respectively. Then, the spacial wave function of the considered tetraquark systems can be written as 
\begin{equation}
 \Psi_L = \sum_{\alpha} A_{\alpha} \left\{ \left[ \psi_{l_1}({\bm{\xi}}_1) \, \psi_{l_2}({\bm{\xi}_2})\right]_{l_{12}} \psi_{l_3}({\bm{\xi}}_3) \right\}_L,    \label{eq:spacial}
\end{equation}
with
\begin{equation}
  \psi_{l_a}({\bm{\xi}}_a) = \left[\frac{2^{l_a+2}\left(2 \nu_a\right)^{l_a+\frac{3}{2}}}{\sqrt{\pi}(2 l_a+1) ! !}\right]^{1/2}{\bm{\xi}}_a^{l_a} \, e^{-\nu_a {\bm{\xi}}_a^2}, 
\end{equation}
where $\alpha = \{l_1,l_2,l_{12},l_3\}$. The expansion coefficient $A_\alpha$ and the Gaussian width parameter $\nu_a$ will be determined by the variational method. For $S$-wave tetraquark systems, one has $l_1=l_2=l_{12}=l_3=L=0$. For $P$-wave tetraquark systems, the lowest energy states will have one of the $l_1$, $l_2$ and $l_3$ being $1$ while the other two being $0$. When $l_1 = 1$ or $l_2 = 1$, the spacial wave function will be denoted as $\rho$-mode excitation where $cc$ or $\bar{q}\bar{q}'$ is antisymmetric in spacial space, and when $l_3 = 1$, the spacial wave function will be denoted as $\lambda$-mode excitation where both $cc$ and $\bar{q}\bar{q}'$ are symmetric in spacial space. 

With the individual wave functions in color, flavor, spin, and spacial spaces given above, one can construct the total wave functions of the $cc\bar{q}\bar{q}'$ $(\bar{q},\bar{q}' = \bar{u}, \bar{d}, \bar{s})$ systems straightforwardly. All configurations considered in the present work are listed in Table~\ref{tab:wave function}. There, the superscript and subscript represent the color SU(3) representation and the spin, respectively. The curly brace $\{~\}$ and square bracket $[~]$ outside $cc$ and $\bar{q}\bar{q}'$ represent the two quarks or antiquarks are symmetric or antisymmetric in flavor space, respectively.

\section{Results and discussion}\label{sec:Results and discussion}

\begin{table*}[htb]
\caption{Masses and eigenvectors for $S$-wave $cc\bar{n}\bar{n}'$, $cc\bar{n}\bar{s}$, and $cc\bar{s}\bar{s}$ $(\bar{n}, \bar{n}' = \bar{u}, \bar{d})$ systems.}
\begin{tabular*}{\textwidth}{@{\extracolsep{\fill}}ccccc}
\hline\hline
 $IJ^P$ & Configuration & $\langle H \rangle$ & Mass (MeV) & Eigenvector \\
\hline
$10^+$ & $ \big(\{cc\}_0^{6_c}\{\bar{n} \bar{n}'\}_0^{\bar{6}_c}\big)_{0}^{1_c} $ &
\multirow{2}{*}{ $ \left[ \begin{array}{rr} 4194 & -81 \\ -81 & 4107 \end{array} \right] $ } &
\multirow{2}{*}{ $ \left[ \begin{array}{c} 4242 \\ 4059  \end{array} \right] $ } &
\multirow{2}{*}{ $ \left[ \begin{array}{rr} -0.86 & -0.51 \\ 0.51 & -0.86 \end{array} \right] $ } \\ 
& $ \big(\{cc\}_1^{\bar{3}_c}\{\bar{n} \bar{n}'\}_1^{3_c}\big)_{0}^{1_c} $ & & & \\[6pt]
$01^+$ & $ \big(\{cc\}_0^{6_c}[\bar{n} \bar{n}']_1^{\bar{6}_c}\big)_{1}^{1_c} $ & 
\multirow{2}{*}{ $ \left[ \begin{array}{rr} 4135 & 51 \\ 51 & 3889 \end{array} \right] $ } &
\multirow{2}{*}{ $ \left[ \begin{array}{c} 4145 \\ 3879  \end{array} \right] $ } &
\multirow{2}{*}{ $ \left[ \begin{array}{rr} -0.98 & 0.20  \\ -0.20 & -0.98 \end{array} \right] $ }  \\ 
& $ \big(\{cc\}_1^{\bar{3}_c}[\bar{n} \bar{n}']_0^{3_c}\big)_{1}^{1_c} $ & & & \\[6pt]
$11^+$ & $\big(\{cc\}_1^{\bar{3}_c}\{\bar{n} \bar{n}'\}_1^{3_c}\big)_{1}^{1_c}$ & $4131$ & $4131$ & $1.00$ \\[6pt]
$12^+$ & $\big(\{cc\}_1^{\bar{3}_c}\{\bar{n} \bar{n}'\}_1^{3_c}\big)_{2}^{1_c}$ & $4176$ & $4176$ & $1.00$ \\[6pt]
$\frac{1}{2}0^+$ & $ \big(\{cc\}_0^{6_c}\{\bar{n} \bar{s}\}_0^{\bar{6}_c}\big)_{0}^{1_c} $ &
\multirow{2}{*}{ $ \left[ \begin{array}{rr} 4285 & -84 \\ -84 & 4230 \end{array} \right] $ } &
\multirow{2}{*}{ $ \left[ \begin{array}{c} 4346 \\ 4169  \end{array} \right] $ } &
\multirow{2}{*}{ $ \left[ \begin{array}{rr} -0.81 & -0.59 \\ 0.59 & -0.81 \end{array} \right] $ } \\ 
& $ \big(\{cc\}_1^{\bar{3}_c}\{\bar{n} \bar{s}\}_1^{3_c}\big)_{0}^{1_c} $ & & & \\[6pt]
$\frac{1}{2}1^+$ & $ \big(\{cc\}_0^{6_c}[\bar{n} \bar{s}]_1^{\bar{6}_c}\big)_{1}^{1_c} $ &
\multirow{2}{*}{ $ \left[ \begin{array}{rr} 4245 & 52 \\ 52 & 4110 \end{array} \right] $ } &
\multirow{2}{*}{ $ \left[ \begin{array}{c} 4263 \\ 4092  \end{array} \right] $ } &
\multirow{2}{*}{ $ \left[ \begin{array}{rr}  -0.95 & 0.32 \\  -0.32 & -0.95 \end{array} \right] $ } \\ 
& $ \big(\{cc\}_1^{\bar{3}_c}[\bar{n} \bar{s}]_0^{3_c}\big)_{1}^{1_c} $ & & & \\[6pt]
$\frac{1}{2}1^+$ & $\big(\{cc\}_1^{\bar{3}_c}\{\bar{n} \bar{s}\}_1^{3_c}\big)_{1}^{1_c}$ & $4255$ & $4255$ & $1.00$ \\[6pt]
$\frac{1}{2}2^+$ & $\big(\{cc\}_1^{\bar{3}_c}\{\bar{n} \bar{s}\}_1^{3_c}\big)_{2}^{1_c}$ & $4301$ & $4301$ & $1.00$ \\[6pt]
$00^+$ & $ \big(\{cc\}_1^{\bar{3}_c}\{\bar{s} \bar{s}\}_1^{3_c}\big)_{0}^{1_c} $ &
\multirow{2}{*}{ $ \left[ \begin{array}{rr} 4370 & -86 \\ -86 & 4347 \end{array} \right] $ } &
\multirow{2}{*}{ $ \left[ \begin{array}{c} 4445 \\ 4272  \end{array} \right] $ } &
\multirow{2}{*}{ $ \left[ \begin{array}{rr} -0.75 & -0.66 \\ 0.66 & -0.75 \end{array} \right] $ } \\ 
& $ \big(\{cc\}_0^{6_c}\{\bar{s} \bar{s}\}_0^{\bar{6}_c}\big)_{0}^{1_c} $ & & & \\[6pt]
$01^+$ & $\big(\{cc\}_1^{\bar{3}_c}\{\bar{s} \bar{s}\}_1^{3_c}\big)_{1}^{1_c}$ & $4374$ & $4374$ & $1.00$ \\[6pt]
$02^+$ & $\big(\{cc\}_1^{\bar{3}_c}\{\bar{s} \bar{s}\}_1^{3_c}\big)_{2}^{1_c}$ & $4423$ & $4423$ & $1.00$ \\
\hline\hline
\end{tabular*}      \label{tab:S_ccqq}
\end{table*}

\begin{table*}[htb]
\caption{Individual contributions of the Hamiltonian to the $cc\bar{n}\bar{n}'$ $(\bar{n},\bar{n}' = \bar{u},\bar{d})$ systems.}
\begin{tabular*}{\textwidth}{@{\extracolsep{\fill}}ccccccccccc}
\hline\hline
  $IJ^P$ & Configuration & $\langle H \rangle$ & $\langle T \rangle$ & $\langle V^{\text{conf}} \rangle$ & $\langle V^{\text{coul}} \rangle$ & $\langle V^{\text{CM}} \rangle$ & $\langle V^{\sigma} \rangle$ & $\langle V^{\pi} \rangle$ & $\langle V^{\text{ten}} \rangle$ & $\langle V^{\text{ls}} \rangle$  \\
\hline
   $01^+$ & $\big(\{cc\}_0^{6_c}[\bar{n} \bar{n}']_1^{\bar{6}_c}\big)_{S=1}^{1_c} $  & $4135$  & $713$ & $193$ & $-618$ & $190$ & $17$ & $25$  & $0$ & $0$ \\
   & $\big(\{cc\}_1^{\bar{3}_c}[\bar{n} \bar{n}']_0^{3_c}\big)_{S=1}^{1_c}$  & $3889$ & $1015$  & $197$ & $-774$  & $1$  & $31$ & $-194$  & $0$ & $0$  \\[6pt]
  $10^+$ & $\big(\{cc\}_0^{6_c}\{\bar{n}\bar{n}'\}_0^{\bar{6}_c}\big)_{S=0}^{1_c}$ & $4194$  &  $675$ & $234$ & $-602$  &  $256$ & $-34$ & $39$ & $0$ &  $0$  \\
    & $\big(\{cc\}_1^{\bar{3}_c}\{\bar{n}\bar{n}'\}_1^{3_c})_{S=0}^{1_c}$  & $4107$ & $723$  & $295$ & $-653$  &  $173$ &  $-40$  & $-16$ & $0$ &  $0$  \\[6pt]
  $00^-$ & $\big(\{cc,\rho\}_1^{6_c}{[\bar{n}\bar{n}']}_1^{\bar{6}_c}\big)^{1_c}_{S=1}$ &  $4167$   & $854$  &  $192$ & $-605$  & $106$  &  $6$  &  $28$  &  $-13$  &  $-27$ \\
  & $\big({\{cc\}}_1^{\bar{3}_c}{[\bar{n} \bar{n}']}_0^{3_c},\lambda\big)^{1_c}_{S=1}$  & $4150$ & $1159$ & $338$  &  $-699$  & $-59$  & $17$  &  $-196$ & $0$ & $-35$ \\[6pt]
  $10^-$ & $\big({\{cc\}}_1^{\bar{3}_c}{\{\bar{n}\bar{n}'\}}_1^{3_c},\lambda\big)^{1_c}_{S=1}$ & $4393$ & $817$  & $456$ & $-581$ & $158$ & $-39$ & $-15$ & $-9$ & $-20$  \\
  & $\big(\{cc,\rho\}^{6_c}_1{\{\bar{n} \bar{n}'\}}^{\bar{6}_c}_0\big)^{1_c}_{S=1} $ & $4299$ & $749$ & $287$ & $-569$ & $220$ & $-35$ & $39$  & $1$ & $-20$  \\
\hline\hline
\end{tabular*}  \label{tab:each_term_V}
\end{table*}

\begin{figure}[htbp]
\subfigure[]{\includegraphics[width=0.5\textwidth]{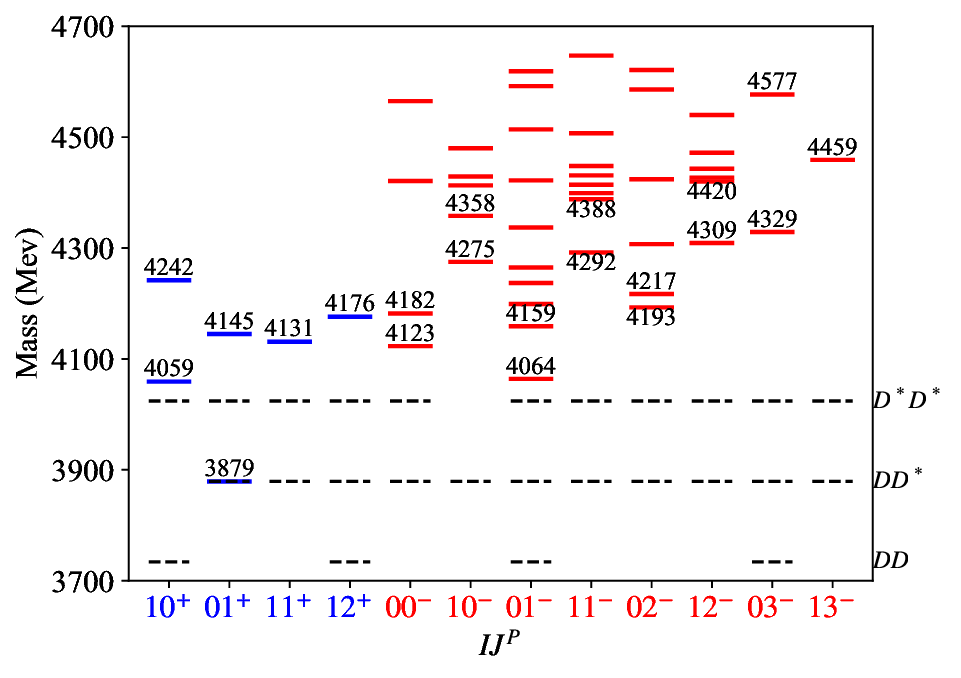} \label{SFig:ccnn}}
\subfigure[]{\includegraphics[width=0.5\textwidth]{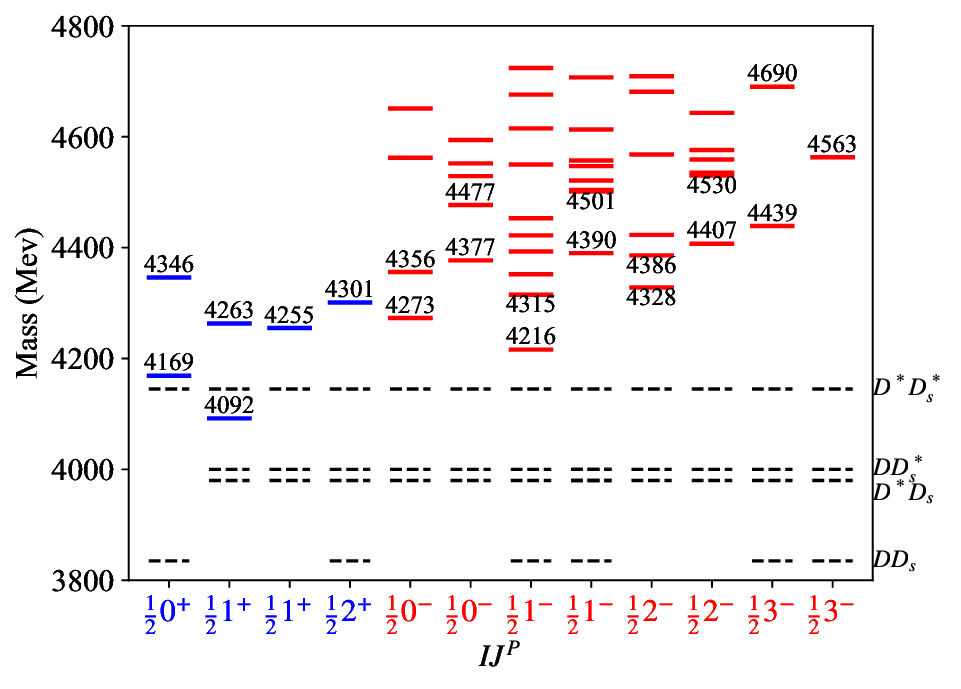} \label{SFig:ccns}}
\subfigure[]{\includegraphics[width=0.5\textwidth]{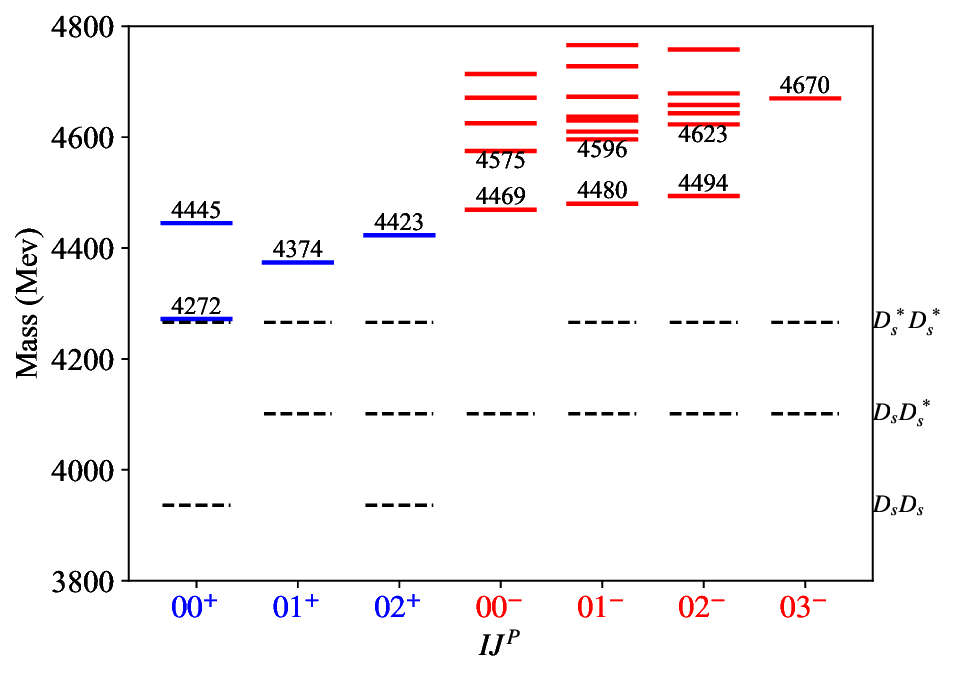} \label{SFig:ccss}}
\caption{The mass spectra of $S$- and $P$-wave $cc\bar{q}\bar{q}'$ $(\bar{q},\bar{q}' = \bar{u}, \bar{d}, \bar{s})$ tetraquark states. Subfigures (a)-(c) correspond to the mass spectra of $cc\bar{n}\bar{n}'$, $cc\bar{n}\bar{s}$, and $cc\bar{s}\bar{s}$ $(\bar{n}, \bar{n}' = \bar{u},\bar{d})$ states, respectively. For $P$-wave states, only those with lower masses are shown for the sake of simplicity.}  \label{SFig}
\end{figure}

\begin{table*}[htbp]
\caption{Root mean square radius (in fm) of the $cc\bar{u}\bar{d}$ state with $IJ^P=01^+$ and mass $3879$ MeV.}
\begin{tabular*}{\textwidth}{@{\extracolsep{\fill}}lcccccccc}
\hline\hline
  $IJ^P$  & Mass  & $\sqrt{\langle r_{12}^2\rangle}$ & $\sqrt{\langle r_{34}^2\rangle}$  & $\sqrt{\langle r_{13}^2\rangle}$ & $\sqrt{\langle r_{24}^2\rangle}$ & $\sqrt{\langle r_{14}^2\rangle}$  & $\sqrt{\langle r_{23}^2\rangle}$ & $\sqrt{{\sum_{i=1}^4 \langle ({\vec{r}}_i-{\vec{R}_{\rm cm}})^2 \rangle}/4}$  \\
\hline
    $01^+$ & $3879$ & $0.54$ & $0.70$ & $0.76$ & $0.76$ & $0.76$ & $0.76$ & $0.48$ \\
\hline\hline
\end{tabular*}   \label{tab:msr_s}
\end{table*}

\begin{figure}[tbp]
\centering
\includegraphics[width=0.45\textwidth]{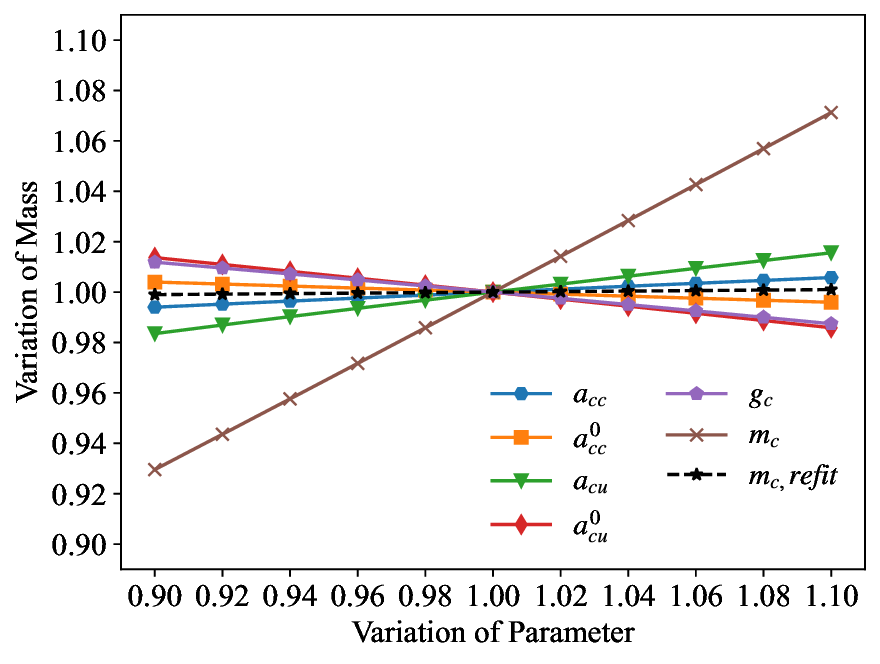}
\caption{Dependence of the mass for the lowest $cc{\bar u}{\bar d}$ state with $IJ^P=01^+$ on parameter values. The vertical and horizontal axes indicate the multiples of the mass and parameter values, respectively, compared to their original values. }
\label{fig:var}
\end{figure}

The reliability of the employed (anti)quark-(anti)quark interactions is crucial for a credible understanding of the multiquark states in a quark model. In the present work, we study the $S$- and $P$-wave $cc\bar{q}\bar{q}'$ $(\bar{q}, \bar{q}' = \bar{u}, \bar{d}, \bar{s})$ tetraquark states in a chiral SU(3) quark model, where the Hamiltonian incorporates the kinetic energies, the OGE potential, the phenomenological confinement potential, and the OBE potential stemming from the couplings of light quarks and chiral fields. The interactions between light antiquarks $\bar{q}$ and $\bar{q}'$ $(\bar{q}, \bar{q}' = \bar{u}, \bar{d}, \bar{s})$ are taken from our previous work of Ref.~\cite{Huang:2018rpb}, which has been shown to be quite successful in reproducing the energies of octet and decuplet baryon ground states, the binding energy of deuteron, and the $NN$ scattering phase shifts and mixing parameters up to total angular momentum $J=6$ in a rather consistent manner. The interactions associated with charm quark, consisting of the OGE potential and the confinement potential, are determined by a good fit to the masses of known charmed mesons and baryons which are listed in Table~\ref{tab:fit}. We choose the Gaussian functions as the trial wave functions in coordinate space and combine them with the color, flavor, and spin wave functions to get the total wave functions of the $S$- and $P$-wave $cc\bar{q}\bar{q}'$ $(\bar{q}, \bar{q}' = \bar{u}, \bar{d}, \bar{s})$ systems that satisfy the Pauli principle. 

Equipped with the Hamiltonian and the trial wave functions, we solve the Schr\"odinger equation by use of the variational method to get the masses and the corresponding eigenvectors for the $S$- and $P$-wave $cc\bar{q}\bar{q}'$ $(\bar{q}, \bar{q}' = \bar{u}, \bar{d}, \bar{s})$ states. The results for $S$-wave $cc\bar{n}\bar{n}'$, $cc\bar{n}\bar{s}$, and $cc\bar{s}\bar{s}$ $(\bar{n}, \bar{n}' = \bar{u}, \bar{d})$ systems are listed in Table~\ref{tab:S_ccqq}. One sees from this table that the lowest $S$-wave states for $cc\bar{n}\bar{n}'$, $cc\bar{n}\bar{s}$, and $cc\bar{s}\bar{s}$ configurations have masses of $3879$ MeV, $4092$ MeV, and $4272$ MeV, respectively, and have quantum numbers of isospin and spin-parity $IJ^P=01^+$, $\frac{1}{2}1^+$, and $00^+$, respectively. The effects of channel couplings of the states with the same quantum numbers are considerable in reducing the energies of these states. Note that there are three $IJ^P=\frac{1}{2}1^+$ states for $cc\bar{n}\bar{s}$ as listed in Table~\ref{tab:wave function}. Two of them have symmetric flavor wave functions for $\bar{n}\bar{s}$, and they do not couple to the other state that has antisymmetric flavor wave functions for $\bar{n}\bar{s}$.

In Table~\ref{tab:each_term_V}, we show the individual contributions of the Hamiltonian to the $cc\bar{n}\bar{n}'$ $(\bar{n},\bar{n}' = \bar{u},\bar{d})$ systems with quantum numbers $IJ^P=01^+$, $10^+$, $00^-$, and $10^-$ (only the lowest two states are listed for negative parity). One sees that for all these states, apart from the kinetic energies, the dominant contributions are coming from the confinement potential and the color coulomb interaction of the OGE potential, followed by the color magnetic interaction of the OGE potential. Significant contributions are also seen from the OBE potential, which reduces the energies of the $cc\bar{n}\bar{n}'$ states with $IJ^P=01^+$, $10^+$, and $00^-$ by $163$ MeV, $56$ MeV, and $179$ MeV, respectively. Note that the contributions of $\pi$ exchange are proportional to the matrix elements of the operator $\langle \boldsymbol{\sigma}_i \cdot \boldsymbol{\sigma}_j \, \boldsymbol{\tau}_i \cdot \boldsymbol{\tau}_j \rangle$ where $\tau^a = \lambda^a$ ($a=1,2,3$). For the states $\big(\{cc\}_1^{\bar{3}_c}[\bar{n} \bar{n}']_0^{3_c}\big)_{S=1}^{1_c}$ with $IJ^P=01^+$ and $\big({\{cc\}}_1^{\bar{3}_c}{[\bar{n} \bar{n}']}_0^{3_c},\lambda\big)^{1_c}_{S=1}$ with $IJ^P=00^-$, one has $\langle \boldsymbol{\sigma}_i \cdot \boldsymbol{\sigma}_j \, \boldsymbol{\tau}_i \cdot \boldsymbol{\tau}_j \rangle = 9$, much bigger than $1$ or $-3$ in other states, as here both isospin and spin of $\bar{n} \bar{n}'$ are zero, which results in the maximum attractions of $\pi$ exchange, reducing the energies of these two states by $194$ MeV and $196$ MeV, respectively. The kinetic energies of these two states are observed to be bigger than the other states. This is because that the strong attractions offered by $\pi$ exchange in these two states compress the spacial sizes of the light quarks in these two states while the kinetic energies are in inverse proportion to the distances of quarks. The contributions from the tensor force and spin-orbit force, marked as $\langle V^{\text{ten}} \rangle$ and $\langle V^{\text{ls}} \rangle$ in the last two columns, are seen to be rather small. This means that the mixing between states with the same quantum numbers $IJ^P$ but different excitation modes in orbit space (excitations of the relative motion between $cc$, the relative motion between $\bar{q}\bar{q}'$, and the relative motion between $cc$ and $\bar{q}\bar{q}'$) are not considerable and the physical states may be approximated to have pure excitation modes.

In Fig.~\ref{SFig}, we present the calculated mass spectra of $S$-wave and $P$-wave $cc\bar{q}\bar{q}'$ $(\bar{q},\bar{q}' = \bar{u},\bar{d},\bar{s})$ states.  For $P$-wave states, only those with lower masses are shown for the sake of simplicity. The meson-meson thresholds of the possible fall-apart decay channels are marked with dashed lines. One sees from Fig.~\ref{SFig:ccnn} that the $cc\bar{n}\bar{n}'$ $(\bar{n},\bar{n}' = \bar{u},\bar{d})$ state with $IJ^P = 01^+$ just locates at $3879$ MeV, very close to the calculated $DD^*$ threshold. This state cannot decay to $DD$, and thus is expected to have a rather small decay width. Experimentally, a narrow state $T_{cc}^+(3875)$ with quantum numbers $IJ^P = 01^+$ and quark constituent $cc\bar{u}\bar{d}$ has been reported by the LHCb Collaboration \cite{LHCb:2021vvq,LHCb:2021auc}. The state with mass $3879$ MeV predicted in our model has the same quantum numbers and quark constituent with the experimentally observed $T_{cc}^+(3875)$. In Table~\ref{tab:msr_s}, we present the root mean square radius of this state calculated in our chiral SU(3) quark model. One sees that the root mean square radius of this state, $\sqrt{{\sum_{i=1}^4 \langle ({\vec{r}}_i-{\vec{R}_{\rm cm}})^2 \rangle}/4}$, is only $0.48$ fm, and moreover, the distances of any pairs of quarks (antiquarks) are less than or equal to $0.76$ fm. This indicates that this state has a rather compact structure. All these demonstrates that the experimentally observed $T_{cc}^+(3875)$ state can be accommodated as a stable and compact $cc\bar{u}\bar{d}$ tetraquark sate in the chiral SU(3) quark model. This state has $4\%$ of the $\big(\{cc\}_0^{6_c}[\bar{u} \bar{d}]_1^{\bar{6}_c}\big)_{1}^{1_c}$ configuration and $96\%$ of the $\big(\{cc\}_1^{\bar{3}_c}[\bar{u} \bar{d}]_0^{3_c}\big)_{1}^{1_c}$ configuration, as can be seen from the eigenvectors listed in Table~\ref{tab:S_ccqq}.

One also sees from Figs.~\ref{SFig:ccnn}-\ref{SFig:ccss} that all the other $S$- and $P$-wave $cc\bar{q}\bar{q}'$ $(\bar{q},\bar{q}' = \bar{u}, \bar{d}, \bar{s})$ states have masses about one hundred to few hundreds MeV higher than the corresponding meson-meson thresholds. For the state $\big({\{cc\}}_1^{\bar{3}_c}{[\bar{n} \bar{n}']}_0^{3_c},\lambda\big)^{1_c}_{S=1}$ with $IJ^P=00^-$, although the $\pi$ exchange offers strong attractions with similar strength to the lowest state $\big(\{cc\}_1^{\bar{3}_c}[\bar{n} \bar{n}']_0^{3_c}\big)_{S=1}^{1_c}$ with $IJ^P=01^+$, the color-coulomb force offers much less attraction and the confinement potential offers much bigger repulsion. Finally this state has a mass more than two hundred MeV higher than the lowest state and the threshold of $DD^\ast$. As these states locate much higher than the corresponding meson-meson thresholds, they are expected to decay easily into two mesons via fall-apart mechanism. Thus, these states are not suggested to be candidates of stable and compact tetraquark states in the chiral SU(3) quark model.
 
In order to see to which extent do our calculated masses of the $cc{\bar q}{\bar q}'$ (${\bar q},{\bar q}'={\bar u},{\bar d}$) tetraquark states depend on the values of the model parameters, in Fig.~\ref{fig:var}, we illustrate the dependence of the mass for the lowest $cc{\bar u}{\bar d}$ state with $IJ^P=01^+$ on the parameter values. Here the vertical and horizontal axes indicate the multiples of the mass and parameter values, respectively, compared to their original values. It is observed that when the value of each of the parameter of $g_c$, $a_{cu}$, $a_{cc}$, $a^0_{cu}$, and $a^0_{cc}$ varies by $10\%$, the mass of the lowest $cc{\bar u}{\bar d}$ state with $IJ^P=01^+$ changes less than $2\%$. This demonstrates that our calculated results are relatively stable against the values of these parameters. Nevertheless, when the charm quark mass, $m_c$, varies by $10\%$, the mass of the lowest $cc{\bar u}{\bar d}$ state with $IJ^P=01^+$ fluctuates by about $7\%$. This can be understood if we notice that $2m_c$ makes up the bulk of the mass of the $cc{\bar u}{\bar d}$ states. In practical calculation, one needs to refit the other parameters when a different value of $m_c$ is chosen. By doing so, the resulting mass of the $cc{\bar u}{\bar d}$ state with $IJ^P=01^+$ keeps nearly unchanged as depicted in Fig.~\ref{fig:var} with a dashed line.

\section{Summary}\label{sec:Summary}

In this work, we systematically explore the mass spectra of the $S$- and $P$-wave $cc\bar{q}\bar{q}'$ $(\bar{q},\bar{q}' = \bar{u},\bar{d},\bar{s})$ systems in a chiral SU(3) quark model. The Hamiltonian consists of the kinetic energy, the confinement potential, the OGE potential, and the OBE potential stemming from the coupling of light quarks (antiquarks) and chiral fields. The interactions between the light antiquarks $\bar{q}$ and $\bar{q}'$ $(\bar{q},\bar{q}' = \bar{u},\bar{d},\bar{s})$ are taken from our previous work of Ref.~\cite{Huang:2018rpb}, that reproduced the energies of octet and decuplet baryon ground states, the binding energy of deuteron, and the $NN$ scattering phase shifts and mixing parameters for total angular momentum up to $J=6$ quite well in a rather consistent manner. The interactions associated with charm quarks are fixed by fitting the masses of known charmed baryons and mesons. The energies and eigenvectors of the $S$- and $P$-wave $cc\bar{q}\bar{q}'$ $(\bar{q},\bar{q}' = \bar{u},\bar{d},\bar{s})$ states are obtained by solving the Schr\"odinger equation via the variational method.

The results show that the lowest $S$-wave $cc\bar{u}\bar{d}$ state with quantum numbers $IJ^P=01^+$ has a mass $3879$ MeV, approximately at the calculated threshold of $DD^\ast$. This state has $4\%$ of the $\big(\{cc\}_0^{6_c}[\bar{u} \bar{d}]_1^{\bar{6}_c}\big)_{1}^{1_c}$ configuration and $96\%$ of the $\big(\{cc\}_1^{\bar{3}_c}[\bar{u} \bar{d}]_0^{3_c}\big)_{1}^{1_c}$ configuration. It cannot decay to the $DD$ channel due to its quantum numbers, and thus is expected to have a rather small decay width. All the mass, width, and quantum numbers of this state are consistent with those of the narrow state $T_{cc}^+(3875)$ reported by the LHCb Collaboration \cite{LHCb:2021vvq,LHCb:2021auc}. Moreover, the calculated root mean square radius of this state is $\sqrt{{\sum_{i=1}^4 \langle ({\vec{r}}_i-{\vec{R}_{\rm cm}})^2 \rangle}/4} = 0.48$ fm, and the distances of any pair of quarks (antiquarks) are less than or equal to $0.76$ fm. All these demonstrate that the experimentally observed $T_{cc}^+(3875)$ state can be accommodated as a compact $cc\bar{u}\bar{d}$ tetraquark state in the chiral SU(3) quark model. 

The energies of all other $S$- and $P$-wave $cc\bar{q}\bar{q}'$ $(\bar{q},\bar{q}' = \bar{u},\bar{d},\bar{s})$ states are about one hundred to several hundreds MeV higher than the thresholds of the corresponding meson-meson channels that can be decayed into via fall-apart mechanism. Thus, they are expected to have large decay widths and not suggested as candidates of compact tetraquark states in the chiral SU(3) quark model.

\begin{acknowledgments}
This work is partially supported by the National Natural Science Foundation of China under Grants No.~12175240 and No.~11635009, and the Fundamental Research Funds for the Central Universities.
\end{acknowledgments}

\bibliographystyle{apsrev}
\bibliography{P_QQqq_ref.bib}

\begin{thebibliography}{31}
\expandafter\ifx\csname natexlab\endcsname\relax\def\natexlab#1{#1}\fi
\expandafter\ifx\csname bibnamefont\endcsname\relax
  \def\bibnamefont#1{#1}\fi
\expandafter\ifx\csname bibfnamefont\endcsname\relax
  \def\bibfnamefont#1{#1}\fi
\expandafter\ifx\csname citenamefont\endcsname\relax
  \def\citenamefont#1{#1}\fi
\expandafter\ifx\csname url\endcsname\relax
  \def\url#1{\texttt{#1}}\fi
\expandafter\ifx\csname urlprefix\endcsname\relax\def\urlprefix{URL }\fi
\providecommand{\bibinfo}[2]{#2}
\providecommand{\eprint}[2][]{\url{#2}}

\bibitem[{\citenamefont{Acosta et~al.}(2004)}]{CDF:2003cab}
\bibinfo{author}{\bibfnamefont{D.}~\bibnamefont{Acosta}} \bibnamefont{et~al.} (\bibinfo{collaboration}{CDF}), \bibinfo{journal}{Phys. Rev. Lett.} \textbf{\bibinfo{volume}{93}}, \bibinfo{pages}{072001} (\bibinfo{year}{2004}).

\bibitem[{\citenamefont{Hosaka et~al.}(2016)\citenamefont{Hosaka, Iijima, Miyabayashi, Sakai, and Yasui}}]{Hosaka:2016pey}
\bibinfo{author}{\bibfnamefont{A.}~\bibnamefont{Hosaka}}, \bibinfo{author}{\bibfnamefont{T.}~\bibnamefont{Iijima}}, \bibinfo{author}{\bibfnamefont{K.}~\bibnamefont{Miyabayashi}}, \bibinfo{author}{\bibfnamefont{Y.}~\bibnamefont{Sakai}}, \bibnamefont{and} \bibinfo{author}{\bibfnamefont{S.}~\bibnamefont{Yasui}}, \bibinfo{journal}{PTEP} \textbf{\bibinfo{volume}{2016}}, \bibinfo{pages}{062C01} (\bibinfo{year}{2016}).

\bibitem[{\citenamefont{Ali et~al.}(2017)\citenamefont{Ali, Lange, and Stone}}]{Ali:2017jda}
\bibinfo{author}{\bibfnamefont{A.}~\bibnamefont{Ali}}, \bibinfo{author}{\bibfnamefont{J.~S.} \bibnamefont{Lange}}, \bibnamefont{and} \bibinfo{author}{\bibfnamefont{S.}~\bibnamefont{Stone}}, \bibinfo{journal}{Prog. Part. Nucl. Phys.} \textbf{\bibinfo{volume}{97}}, \bibinfo{pages}{123} (\bibinfo{year}{2017}).

\bibitem[{\citenamefont{Guo et~al.}(2018)\citenamefont{Guo, Hanhart, Mei\ss{}ner, Wang, Zhao, and Zou}}]{Guo:2017jvc}
\bibinfo{author}{\bibfnamefont{F.-K.} \bibnamefont{Guo}}, \bibinfo{author}{\bibfnamefont{C.}~\bibnamefont{Hanhart}}, \bibinfo{author}{\bibfnamefont{U.-G.} \bibnamefont{Mei\ss{}ner}}, \bibinfo{author}{\bibfnamefont{Q.}~\bibnamefont{Wang}}, \bibinfo{author}{\bibfnamefont{Q.}~\bibnamefont{Zhao}}, \bibnamefont{and} \bibinfo{author}{\bibfnamefont{B.-S.} \bibnamefont{Zou}}, \bibinfo{journal}{Rev. Mod. Phys.} \textbf{\bibinfo{volume}{90}}, \bibinfo{pages}{015004} (\bibinfo{year}{2018}), \bibinfo{note}{[Erratum: Rev. Mod. Phys. 94, 029901 (2022)]}.

\bibitem[{\citenamefont{Esposito et~al.}(2017)\citenamefont{Esposito, Pilloni, and Polosa}}]{Esposito:2016noz}
\bibinfo{author}{\bibfnamefont{A.}~\bibnamefont{Esposito}}, \bibinfo{author}{\bibfnamefont{A.}~\bibnamefont{Pilloni}}, \bibnamefont{and} \bibinfo{author}{\bibfnamefont{A.~D.} \bibnamefont{Polosa}}, \bibinfo{journal}{Phys. Rept.} \textbf{\bibinfo{volume}{668}}, \bibinfo{pages}{1} (\bibinfo{year}{2017}).

\bibitem[{\citenamefont{Lebed et~al.}(2017)\citenamefont{Lebed, Mitchell, and Swanson}}]{Lebed:2016hpi}
\bibinfo{author}{\bibfnamefont{R.~F.} \bibnamefont{Lebed}}, \bibinfo{author}{\bibfnamefont{R.~E.} \bibnamefont{Mitchell}}, \bibnamefont{and} \bibinfo{author}{\bibfnamefont{E.~S.} \bibnamefont{Swanson}}, \bibinfo{journal}{Prog. Part. Nucl. Phys.} \textbf{\bibinfo{volume}{93}}, \bibinfo{pages}{143} (\bibinfo{year}{2017}).

\bibitem[{\citenamefont{Richard}(2016)}]{Richard:2016eis}
\bibinfo{author}{\bibfnamefont{J.-M.} \bibnamefont{Richard}}, \bibinfo{journal}{Few Body Syst.} \textbf{\bibinfo{volume}{57}}, \bibinfo{pages}{1185} (\bibinfo{year}{2016}).

\bibitem[{\citenamefont{Chen et~al.}(2016)\citenamefont{Chen, Chen, Liu, and Zhu}}]{Chen:2016qju}
\bibinfo{author}{\bibfnamefont{H.-X.} \bibnamefont{Chen}}, \bibinfo{author}{\bibfnamefont{W.}~\bibnamefont{Chen}}, \bibinfo{author}{\bibfnamefont{X.}~\bibnamefont{Liu}}, \bibnamefont{and} \bibinfo{author}{\bibfnamefont{S.-L.} \bibnamefont{Zhu}}, \bibinfo{journal}{Phys. Rept.} \textbf{\bibinfo{volume}{639}}, \bibinfo{pages}{1} (\bibinfo{year}{2016}).

\bibitem[{\citenamefont{Liu et~al.}(2019)\citenamefont{Liu, Chen, Chen, Liu, and Zhu}}]{Liu:2019zoy}
\bibinfo{author}{\bibfnamefont{Y.-R.} \bibnamefont{Liu}}, \bibinfo{author}{\bibfnamefont{H.-X.} \bibnamefont{Chen}}, \bibinfo{author}{\bibfnamefont{W.}~\bibnamefont{Chen}}, \bibinfo{author}{\bibfnamefont{X.}~\bibnamefont{Liu}}, \bibnamefont{and} \bibinfo{author}{\bibfnamefont{S.-L.} \bibnamefont{Zhu}}, \bibinfo{journal}{Prog. Part. Nucl. Phys.} \textbf{\bibinfo{volume}{107}}, \bibinfo{pages}{237} (\bibinfo{year}{2019}).

\bibitem[{\citenamefont{Brambilla et~al.}(2020)\citenamefont{Brambilla, Eidelman, Hanhart, Nefediev, Shen, Thomas, Vairo, and Yuan}}]{Brambilla:2019esw}
\bibinfo{author}{\bibfnamefont{N.}~\bibnamefont{Brambilla}}, \bibinfo{author}{\bibfnamefont{S.}~\bibnamefont{Eidelman}}, \bibinfo{author}{\bibfnamefont{C.}~\bibnamefont{Hanhart}}, \bibinfo{author}{\bibfnamefont{A.}~\bibnamefont{Nefediev}}, \bibinfo{author}{\bibfnamefont{C.-P.} \bibnamefont{Shen}}, \bibinfo{author}{\bibfnamefont{C.~E.} \bibnamefont{Thomas}}, \bibinfo{author}{\bibfnamefont{A.}~\bibnamefont{Vairo}}, \bibnamefont{and} \bibinfo{author}{\bibfnamefont{C.-Z.} \bibnamefont{Yuan}}, \bibinfo{journal}{Phys. Rept.} \textbf{\bibinfo{volume}{873}}, \bibinfo{pages}{1} (\bibinfo{year}{2020}).

\bibitem[{\citenamefont{Aaij et~al.}(2022{\natexlab{a}})}]{LHCb:2021vvq}
\bibinfo{author}{\bibfnamefont{R.}~\bibnamefont{Aaij}} \bibnamefont{et~al.} (\bibinfo{collaboration}{LHCb Collaboration}), \bibinfo{journal}{Nature Phys.} \textbf{\bibinfo{volume}{18}}, \bibinfo{pages}{751} (\bibinfo{year}{2022}{\natexlab{a}}).

\bibitem[{\citenamefont{Aaij et~al.}(2022{\natexlab{b}})}]{LHCb:2021auc}
\bibinfo{author}{\bibfnamefont{R.}~\bibnamefont{Aaij}} \bibnamefont{et~al.} (\bibinfo{collaboration}{LHCb Collaboration}), \bibinfo{journal}{Nature Commun.} \textbf{\bibinfo{volume}{13}}, \bibinfo{pages}{3351} (\bibinfo{year}{2022}{\natexlab{b}}).

\bibitem[{\citenamefont{Ebert et~al.}(2007)\citenamefont{Ebert, Faustov, Galkin, and Lucha}}]{Ebert:2007rn}
\bibinfo{author}{\bibfnamefont{D.}~\bibnamefont{Ebert}}, \bibinfo{author}{\bibfnamefont{R.~N.} \bibnamefont{Faustov}}, \bibinfo{author}{\bibfnamefont{V.~O.} \bibnamefont{Galkin}}, \bibnamefont{and} \bibinfo{author}{\bibfnamefont{W.}~\bibnamefont{Lucha}}, \bibinfo{journal}{Phys. Rev. D} \textbf{\bibinfo{volume}{76}}, \bibinfo{pages}{114015} (\bibinfo{year}{2007}).

\bibitem[{\citenamefont{L\"u et~al.}(2020)\citenamefont{L\"u, Chen, and Dong}}]{Lu:2020rog}
\bibinfo{author}{\bibfnamefont{Q.-F.} \bibnamefont{L\"u}}, \bibinfo{author}{\bibfnamefont{D.-Y.} \bibnamefont{Chen}}, \bibnamefont{and} \bibinfo{author}{\bibfnamefont{Y.-B.} \bibnamefont{Dong}}, \bibinfo{journal}{Phys. Rev. D} \textbf{\bibinfo{volume}{102}}, \bibinfo{pages}{034012} (\bibinfo{year}{2020}).

\bibitem[{\citenamefont{Wang et~al.}(2022)\citenamefont{Wang, Li, An, Deng, and Xie}}]{Wang:2022clw}
\bibinfo{author}{\bibfnamefont{J.-B.} \bibnamefont{Wang}}, \bibinfo{author}{\bibfnamefont{G.}~\bibnamefont{Li}}, \bibinfo{author}{\bibfnamefont{C.-S.} \bibnamefont{An}}, \bibinfo{author}{\bibfnamefont{C.-R.} \bibnamefont{Deng}}, \bibnamefont{and} \bibinfo{author}{\bibfnamefont{J.-J.} \bibnamefont{Xie}}, \bibinfo{journal}{Eur. Phys. J. C} \textbf{\bibinfo{volume}{82}}, \bibinfo{pages}{721} (\bibinfo{year}{2022}).

\bibitem[{\citenamefont{Vijande et~al.}(2009)\citenamefont{Vijande, Valcarce, and Barnea}}]{Vijande:2009kj}
\bibinfo{author}{\bibfnamefont{J.}~\bibnamefont{Vijande}}, \bibinfo{author}{\bibfnamefont{A.}~\bibnamefont{Valcarce}}, \bibnamefont{and} \bibinfo{author}{\bibfnamefont{N.}~\bibnamefont{Barnea}}, \bibinfo{journal}{Phys. Rev. D} \textbf{\bibinfo{volume}{79}}, \bibinfo{pages}{074010} (\bibinfo{year}{2009}).

\bibitem[{\citenamefont{Meng et~al.}(2021)\citenamefont{Meng, Hiyama, Hosaka, Oka, Gubler, Can, Takahashi, and Zong}}]{Meng:2020knc}
\bibinfo{author}{\bibfnamefont{Q.}~\bibnamefont{Meng}}, \bibinfo{author}{\bibfnamefont{E.}~\bibnamefont{Hiyama}}, \bibinfo{author}{\bibfnamefont{A.}~\bibnamefont{Hosaka}}, \bibinfo{author}{\bibfnamefont{M.}~\bibnamefont{Oka}}, \bibinfo{author}{\bibfnamefont{P.}~\bibnamefont{Gubler}}, \bibinfo{author}{\bibfnamefont{K.~U.} \bibnamefont{Can}}, \bibinfo{author}{\bibfnamefont{T.~T.} \bibnamefont{Takahashi}}, \bibnamefont{and} \bibinfo{author}{\bibfnamefont{H.~S.} \bibnamefont{Zong}}, \bibinfo{journal}{Phys. Lett. B} \textbf{\bibinfo{volume}{814}}, \bibinfo{pages}{136095} (\bibinfo{year}{2021}).

\bibitem[{\citenamefont{Deng and Zhu}(2022)}]{Deng:2021gnb}
\bibinfo{author}{\bibfnamefont{C.}~\bibnamefont{Deng}} \bibnamefont{and} \bibinfo{author}{\bibfnamefont{S.-L.} \bibnamefont{Zhu}}, \bibinfo{journal}{Phys. Rev. D} \textbf{\bibinfo{volume}{105}}, \bibinfo{pages}{054015} (\bibinfo{year}{2022}).

\bibitem[{\citenamefont{Navarra et~al.}(2007)\citenamefont{Navarra, Nielsen, and Lee}}]{Navarra:2007yw}
\bibinfo{author}{\bibfnamefont{F.~S.} \bibnamefont{Navarra}}, \bibinfo{author}{\bibfnamefont{M.}~\bibnamefont{Nielsen}}, \bibnamefont{and} \bibinfo{author}{\bibfnamefont{S.~H.} \bibnamefont{Lee}}, \bibinfo{journal}{Phys. Lett. B} \textbf{\bibinfo{volume}{649}}, \bibinfo{pages}{166} (\bibinfo{year}{2007}).

\bibitem[{\citenamefont{Du et~al.}(2013)\citenamefont{Du, Chen, Chen, and Zhu}}]{Du:2012wp}
\bibinfo{author}{\bibfnamefont{M.-L.} \bibnamefont{Du}}, \bibinfo{author}{\bibfnamefont{W.}~\bibnamefont{Chen}}, \bibinfo{author}{\bibfnamefont{X.-L.} \bibnamefont{Chen}}, \bibnamefont{and} \bibinfo{author}{\bibfnamefont{S.-L.} \bibnamefont{Zhu}}, \bibinfo{journal}{Phys. Rev. D} \textbf{\bibinfo{volume}{87}}, \bibinfo{pages}{014003} (\bibinfo{year}{2013}).

\bibitem[{\citenamefont{Agaev et~al.}(2022)\citenamefont{Agaev, Azizi, and Sundu}}]{Agaev:2021vur}
\bibinfo{author}{\bibfnamefont{S.~S.} \bibnamefont{Agaev}}, \bibinfo{author}{\bibfnamefont{K.}~\bibnamefont{Azizi}}, \bibnamefont{and} \bibinfo{author}{\bibfnamefont{H.}~\bibnamefont{Sundu}}, \bibinfo{journal}{Nucl. Phys. B} \textbf{\bibinfo{volume}{975}}, \bibinfo{pages}{115650} (\bibinfo{year}{2022}).

\bibitem[{\citenamefont{Ikeda et~al.}(2014)\citenamefont{Ikeda, Charron, Aoki, Doi, Hatsuda, Inoue, Ishii, Murano, Nemura, and Sasaki}}]{Ikeda:2013vwa}
\bibinfo{author}{\bibfnamefont{Y.}~\bibnamefont{Ikeda}}, \bibinfo{author}{\bibfnamefont{B.}~\bibnamefont{Charron}}, \bibinfo{author}{\bibfnamefont{S.}~\bibnamefont{Aoki}}, \bibinfo{author}{\bibfnamefont{T.}~\bibnamefont{Doi}}, \bibinfo{author}{\bibfnamefont{T.}~\bibnamefont{Hatsuda}}, \bibinfo{author}{\bibfnamefont{T.}~\bibnamefont{Inoue}}, \bibinfo{author}{\bibfnamefont{N.}~\bibnamefont{Ishii}}, \bibinfo{author}{\bibfnamefont{K.}~\bibnamefont{Murano}}, \bibinfo{author}{\bibfnamefont{H.}~\bibnamefont{Nemura}}, \bibnamefont{and} \bibinfo{author}{\bibfnamefont{K.}~\bibnamefont{Sasaki}}, \bibinfo{journal}{Phys. Lett. B} \textbf{\bibinfo{volume}{729}}, \bibinfo{pages}{85} (\bibinfo{year}{2014}).

\bibitem[{\citenamefont{Junnarkar et~al.}(2019)\citenamefont{Junnarkar, Mathur, and Padmanath}}]{Junnarkar:2018twb}
\bibinfo{author}{\bibfnamefont{P.}~\bibnamefont{Junnarkar}}, \bibinfo{author}{\bibfnamefont{N.}~\bibnamefont{Mathur}}, \bibnamefont{and} \bibinfo{author}{\bibfnamefont{M.}~\bibnamefont{Padmanath}}, \bibinfo{journal}{Phys. Rev. D} \textbf{\bibinfo{volume}{99}}, \bibinfo{pages}{034507} (\bibinfo{year}{2019}).

\bibitem[{\citenamefont{Li et~al.}(2013)\citenamefont{Li, Sun, Liu, and Zhu}}]{Li:2012ss}
\bibinfo{author}{\bibfnamefont{N.}~\bibnamefont{Li}}, \bibinfo{author}{\bibfnamefont{Z.-F.} \bibnamefont{Sun}}, \bibinfo{author}{\bibfnamefont{X.}~\bibnamefont{Liu}}, \bibnamefont{and} \bibinfo{author}{\bibfnamefont{S.-L.} \bibnamefont{Zhu}}, \bibinfo{journal}{Phys. Rev. D} \textbf{\bibinfo{volume}{88}}, \bibinfo{pages}{114008} (\bibinfo{year}{2013}).

\bibitem[{\citenamefont{Huang and Wang}(2018)}]{Huang:2018rpb}
\bibinfo{author}{\bibfnamefont{F.}~\bibnamefont{Huang}} \bibnamefont{and} \bibinfo{author}{\bibfnamefont{W.~L.} \bibnamefont{Wang}}, \bibinfo{journal}{Phys. Rev. D} \textbf{\bibinfo{volume}{98}}, \bibinfo{pages}{074018} (\bibinfo{year}{2018}).

\bibitem[{\citenamefont{Zhang et~al.}(1997)\citenamefont{Zhang, Yu, Shen, Dai, Faessler, and Straub}}]{Zhang:1997ny}
\bibinfo{author}{\bibfnamefont{Z.~Y.} \bibnamefont{Zhang}}, \bibinfo{author}{\bibfnamefont{Y.~W.} \bibnamefont{Yu}}, \bibinfo{author}{\bibfnamefont{P.~N.} \bibnamefont{Shen}}, \bibinfo{author}{\bibfnamefont{L.~R.} \bibnamefont{Dai}}, \bibinfo{author}{\bibfnamefont{A.}~\bibnamefont{Faessler}}, \bibnamefont{and} \bibinfo{author}{\bibfnamefont{U.}~\bibnamefont{Straub}}, \bibinfo{journal}{Nucl. Phys. A} \textbf{\bibinfo{volume}{625}}, \bibinfo{pages}{59} (\bibinfo{year}{1997}).

\bibitem[{\citenamefont{Huang and Zhang}(2004)}]{Huang:2004ke}
\bibinfo{author}{\bibfnamefont{F.}~\bibnamefont{Huang}} \bibnamefont{and} \bibinfo{author}{\bibfnamefont{Z.~Y.} \bibnamefont{Zhang}}, \bibinfo{journal}{Phys. Rev. C} \textbf{\bibinfo{volume}{70}}, \bibinfo{pages}{064004} (\bibinfo{year}{2004}).

\bibitem[{\citenamefont{Huang et~al.}(2004{\natexlab{a}})\citenamefont{Huang, Zhang, and Yu}}]{Huang:2004sj}
\bibinfo{author}{\bibfnamefont{F.}~\bibnamefont{Huang}}, \bibinfo{author}{\bibfnamefont{Z.~Y.} \bibnamefont{Zhang}}, \bibnamefont{and} \bibinfo{author}{\bibfnamefont{Y.~W.} \bibnamefont{Yu}}, \bibinfo{journal}{Phys. Rev. C} \textbf{\bibinfo{volume}{70}}, \bibinfo{pages}{044004} (\bibinfo{year}{2004}{\natexlab{a}}).

\bibitem[{\citenamefont{Workman et~al.}(2022)}]{ParticleDataGroup:2022pth}
\bibinfo{author}{\bibfnamefont{R.~L.} \bibnamefont{Workman}} \bibnamefont{et~al.} (\bibinfo{collaboration}{Particle Data Group}), \bibinfo{journal}{PTEP} \textbf{\bibinfo{volume}{2022}}, \bibinfo{pages}{083C01} (\bibinfo{year}{2022}).

\bibitem[{\citenamefont{Huang et~al.}(2004{\natexlab{b}})\citenamefont{Huang, Zhang, Yu, and Zou}}]{Huang:2003we}
\bibinfo{author}{\bibfnamefont{F.}~\bibnamefont{Huang}}, \bibinfo{author}{\bibfnamefont{Z.~Y.} \bibnamefont{Zhang}}, \bibinfo{author}{\bibfnamefont{Y.~W.} \bibnamefont{Yu}}, \bibnamefont{and} \bibinfo{author}{\bibfnamefont{B.~S.} \bibnamefont{Zou}}, \bibinfo{journal}{Phys. Lett. B} \textbf{\bibinfo{volume}{586}}, \bibinfo{pages}{69} (\bibinfo{year}{2004}{\natexlab{b}}).

\bibitem[{\citenamefont{Wang et~al.}(2007)\citenamefont{Wang, Huang, Zhang, Yu, and Liu}}]{Wang:2007kb}
\bibinfo{author}{\bibfnamefont{W.~L.} \bibnamefont{Wang}}, \bibinfo{author}{\bibfnamefont{F.}~\bibnamefont{Huang}}, \bibinfo{author}{\bibfnamefont{Z.~Y.} \bibnamefont{Zhang}}, \bibinfo{author}{\bibfnamefont{Y.~W.} \bibnamefont{Yu}}, \bibnamefont{and} \bibinfo{author}{\bibfnamefont{F.}~\bibnamefont{Liu}}, \bibinfo{journal}{J. Phys. G} \textbf{\bibinfo{volume}{34}}, \bibinfo{pages}{1771} (\bibinfo{year}{2007}).

\end{thebibliography}

\end{document}